\documentclass[a4paper,prd,11pt,noshowkeys,notitlepage,nofootinbib,superscriptaddress]{revtex4}
\pdfoutput=1  %necessary for JHEP
\usepackage[utf8]{inputenc}
\usepackage{ae,aecompl}
\usepackage{amsmath,amssymb,mathrsfs}
\usepackage{graphicx}
\usepackage{ccnishi}
\usepackage{dsfont}
\usepackage{slashed}
\usepackage{units}
\usepackage[normalem]{ulem}

\usepackage[usenames,dvipsnames]{xcolor} % svgnames
\usepackage{hyperref}
\hypersetup{
    colorlinks=true,        % false: boxed links; true: colored links
    linkcolor=purple,       % color of internal links, e.g., eqs.
    citecolor=JungleGreen,  % color of links to bibliography
    filecolor=magenta,      % color of file links
}

\providecommand{\cM}{\mathscr{M}}

\providecommand{\cY}{\mathscr{Y}}
\providecommand{\cO}{\mathcal{O}}

\providecommand{\dst}{\displaystyle}
\providecommand{\tY}{\tilde{Y}}

\providecommand{\pmnssm}{V_{\rm pmns}}

\providecommand{\hY}{\hat{Y}}

\providecommand{\lag}{\mathscr{L}}

\providecommand{\Gammaconv}{\Gamma^{\rm conv}_{\mu \rightarrow e}}
\providecommand{\Gammacapt}[1]{\Gamma^{\rm capt}_{\rm #1}}

\providecommand{\tf}{\tilde{f}}
\renewcommand{\tg}{\tilde{g}}
\providecommand{\tF}{\tilde{F}}

\makeatletter
\newcommand*\bigcdot{\mathpalette\bigcdot@{.5}}
\newcommand*\bigcdot@[2]{\mathbin{\vcenter{\hbox{\scalebox{#2}{$\m@th#1\bullet$}}}}}
\makeatother

\begin{document}
%%%%%%%%%%%%%%%%%%%%%%%%%%%%%%%%%%%%%%%%%%%%%%%%%
\title{
Leptonic CP violation from a vector-like lepton
}
\author{A.~L.~Cherchiglia}
\thanks{E-mail: adriano.cherchiglia@ufabc.edu.br}
\affiliation{Centro de Ci\^{e}ncias Naturais e Humanas,
Universidade Federal do ABC, 09.210-170,
Santo Andr\'{e}-SP, Brasil}
\author{G.~De~Conto}
\thanks{E-mail: george.de.conto@gmail.com}
\affiliation{Centro de Ci\^{e}ncias Naturais e Humanas,
Universidade Federal do ABC, 09.210-170,
Santo Andr\'{e}-SP, Brasil}
\author{C.~C.~Nishi}
\thanks{E-mail: celso.nishi@ufabc.edu.br}
\affiliation{Centro de Matem\'{a}tica, Computa\c{c}\~{a}o e Cogni\c{c}\~{a}o,
Universidade Federal do ABC, 09.210-170,
Santo Andr\'{e}-SP, Brasil}
%\affiliation{Universidade Federal do ABC - UFABC, Santo Andr\'{e}, SP, Brasil}

% \date{\today}
%%%%%%%%%%%%%%%%%%%%%%%%%%%%%%%%%%%%%%%%%%%%%%%%%
\begin{abstract}
Leptonic CP violation is yet to be confirmed as an additional source of CP violation in fundamental interactions.
We study the case where leptonic CP violation is spontaneous and is induced by the mixing with a heavy charged vector-like lepton (VLL).
We show that the non-decoupling of this VLL is linked with the presence of CP violation and its coupling with the SM leptons are partly fixed from the SM Yukawas.
Due to the large leptonic mixing angles, these couplings are typically of the same order and there is no flavor preference.
Strong but not definitive constraints come from charged lepton flavor violating processes because the VLL can decouple from one or two leptonic flavors in very special points of parameter space.
These special points are very sensitive to the neutrino Majorana phases.

\end{abstract}
%%%%%%%%%%%%%%%%%%%%%%%%%%%%%%%%%%%%%%%%%%%%%%%%%
% \pacs{12.60.Fr, 14.80.Cp, 11.30.Qc, 02.20.-a}
%\keywords{ }
%\twocolumn
\maketitle
% \tableofcontents
%%%%%%%%%%%%%%%%%%%%%%%%%%%%%%%%%%%%%%%%%%%%%%%%%
\section{Introduction}
\label{sec:intro}

The discovery that the microscopic physical laws violate CP symmetry dates back to the kaon system\,\cite{cronin} and its origin within the SM is associated to one single phase residing in the complex Yukawa couplings connecting the Higgs with the quarks.
This single source has been extensively tested in B factories\,\cite{babar.belle}
but we have not yet found any other source. 
New sources of CP violation, though, are necessary to explain the asymmetry between matter and antimatter in the universe.
One new source is indeed expected to be residing in the leptonic sector where CP violation should appear in much the same way as for quarks.
Hints of nonzero values for the Dirac CP phase are already present in neutrino oscillation experiments\,\cite{t2k.2020}.
This leptonic CP violation, in turn, may seed the baryon asymmetry of the universe through a dynamically generated lepton asymmetry\,\cite{leptog}.

It may happen that at high energies nature is CP symmetric but as we go to lower energies we see spontaneous CP breaking\,\cite{lee}. This is intimately connected to the Nelson-Barr idea of solving the strong CP problem\,\cite{nelson,barr}.
In this kind of construction, heavy vector-like quarks (VLQs) that mix with SM quarks are necessary to transmit the spontaneously broken CP.
As a consequence, these vector-like quarks of Nelson-Barr type (NB-VLQs) cannot decouple from the SM\,\cite{BBP} and their coupling with the SM generically inherit the hierarchy of the CKM and quark Yukawas\,\cite{nb-vlq,nb-vlq:fit}.
A coincidence of scales\,\cite{vecchi.1} is also generically necessary and this feature may be explained by resorting to strong dynamics\,\cite{vecchi.2}.

In a GUT context, as in the original model of Nelson\,\cite{nelson}, CP violation in the leptonic sector may also be spontaneous (see e.g. Refs.\,\cite{gut:scpv}). In this case, the CP violation is transmitted by vector-like leptons (VLLs) and these VLL of Nelson-Barr type (NB-VLLs) are the focus of this paper.
In analogy to the quark case, the requirement that leptonic CP violation be induced by the mixing between these heavy leptons with the SM leptons will constrain the structure of these mixing terms. Uncovering this structure will be one of the goals here. 
The second goal is to compare it to a generic VLL unrelated to the origin of CP violation.
We will focus on singlet charged VLLs. 
The case of a neutral VLL leads to the usual type I seesaw scenario with CP breaking induced by the complex heavy neutrino Majorana mass term and this has been studied, e.g., in Ref.\,\cite{branco:common}.
Our scenario disentangles the origin of leptonic CP violation from other flavor problems, focusing exclusively on the former. So it differs, for example, from models where the mixing with heavy leptons induce the \emph{whole package}\,\cite{falkowski} ---Yukawas and mixing of SM leptons.

Because of their vector-like nature, the addition of VLLs are also one of the simplest complete SM extensions that deserve to be studied by their own.
Not interacting with gluons, colliders searches put much weaker constraints on VLLs\,\cite{vll.lhc} than on vector-like quarks\,\cite{cms:vlq,atlas:vlq}.
Consequently, low energy probes such as electroweak observables or charged lepton flavor violation (CLFV) are expected to have a much more extended reach to these heavy leptons\,\cite{calibbi:review}.
The latter, of course, requires that the heavy leptons possess flavor violating couplings to the SM leptons.
And indeed, this is what we expect for a charged NB-VLL, thinking in analogy to NB-VLQs.

Another motivation concerns the recent result for the muon magnetic moment \cite{Muong-2:2021ojo}, which shows a tantalizing discrepancy to the standard model of more than $4 \sigma$.\footnote{For the theoretical prediction in the SM we adopt the value of the white paper \cite{Aoyama:2020ynm}. Other groups from lattice \cite{Borsanyi:2020mff,Lehner:2020crt} provide a different prediction, which brings the theoretical and experimental value closer to it other.}
Since the old data \cite{Bennett:2006fi}, it is known that this anomaly has the size of electroweak physics but mediated by some new physics\,\cite{strumia}.
Although vector-like leptons appear as natural candidates, if only one representation is present, the contribution to $(g-2)_{\mu}$ is either negative or not viable (requiring non-perturbative or phenomenological excluded couplings) \cite{freitas}. The situation is different if more than one representation is allowed simultaneously \cite{freitas,2.vll,Crivellin:2018qmi} or if a singlet scalar is added together with a doublet or singlet VLL. The latter is one of the remaining possibilities to explain $(g-2)_{\mu}$ by adding only two new fields to the SM \cite{Athron:2021iuf}.

In this article, we define these VLLs of Nelson-Barr type as messengers of leptonic CP violation to the SM, analyze its structure in detail for the case of one VLL and discuss the phenomenological constraints, primarily focusing on low energy observables.
The outline of this article is as follows: In Sec.\,\ref{sec:NB} we introduce our model of one vector-like lepton of Nelson-Barr type and the employed parametrization is explained in Sec.\,\ref{sec:diag.1}.
We show that typically the vector-like lepton couples to all lepton flavors with similar strength.
The couplings to the $W$ and $Z$ are reviewed in Sec.\,\ref{sec:obs}
while the relevant processes of charged lepton flavor violation are discussed in 
Sec.\,\ref{sec:clv}.
Section \ref{sec:special} presents and proves the existence of special points in parameter space where the vector-like lepton may decouple from one or, more specially, two leptonic flavors. For these special cases, lepton flavor violating processes are suppressed.
We gather the relevant constraints in Sec.\,\ref{sec:params} and show the available parameter space of the model.
The conclusions can be read in Sec.\,\ref{sec:conclu}.

%%%%%%%%%%%%%%%%%%%%%%%%%%%%%%%%%%%%%%%%%%%
\section{Vector-like leptons of Nelson-Barr type}
\label{sec:NB}

Let us define our model of vector-like leptons (VLLs) of Nelson-Barr type. The definition is analogous to the case of vector-like quarks (VLQs) of Nelson-Barr type defined in Ref.\,\cite{nb-vlq} and it is motivated by GUT models that solve the strong CP problem through the Nelson-Barr mechanism, requiring the addition of multiplets that include VLLs in addition to VLQs.
For simplicity, we will denote VLLs of Nelson-Barr type as NB-VLLs.

The original model of Nelson\,\cite{nelson}, a $SU(5)$ GUT, is just such a case where one family of a vector-like $\mathbf{10}$, paired with $\mathbf{10}^*$, and a vector-like $\mathbf{5}^*$, paired with $\mathbf{5}$, are introduced.
The representations $\mathbf{10}$ and $\mathbf{10}^*$ contain the additional vector-like charged leptons, singlets of the SM, while the $\mathbf{5}^*$ and $\mathbf{5}$ contain vector-like doublet leptons.

Motivated by the Nelson model, we will consider the \emph{singlet} NB-VLL model with one VLL $(E_R,E_L)$ of charge $-1$ in addition to the SM.
In the Nelson model, it corresponds to the simplified situation where the additional $\mathbf{5}$ is decoupled from the rest of the fields while the $\mathbf{10}^*$, containing the charged VLL singlet $E_R$, couples to the SM. The additional vector-like quarks in $\mathbf{10}^*$ ---a quark doublet and an up-type singlet--- completes the model to solve the strong CP problem through the Nelson-Barr mechanism.

We will study the consequences of these VLLs with masses at the TeV scale or below while we will assume that the active neutrino masses come from some mechanism acting at much higher scales.
We will also assume that the neutrino mass generation mechanism is disconnected from CP violation\,\footnote{Therefore, if the baryon asymmetry of the universe is explained by leptogenesis, it needs to be low scale. For bariogenesis, there are some constructions which require the inclusion of VLL, for instance \cite{bario}.} 
and that the description through the Weinberg operator is valid with real coefficients. The CP violation in the lepton sector will come entirely from the charged lepton sector.
Without loss of generality, we can choose a basis where the neutrino mass matrix is diagonal
and the leptonic mixing comes entirely from the charged lepton part as well.

We define the singlet NB-VLL model by the addition of one singlet VLL $E_R,E_L$ of charge $-1$. For the definition, it is useful to write the lagrangian for the case of one 
\begin{equation}
\label{lag:E:gen}
\text{Singlet Generic VLL:}\quad
-\lag= \bar{\ell}_{iL} Y^{e}_{ij} H e_{jR} + 
    \bar{\ell}_{iL} Y^{E}_{i} H E_{R} + M_E\bar{E}_{L} E_{R} +h.c.,
\end{equation}
where $\ell_{iL}$ are the SM lepton doublets and $e_{jR}$ are the charged SM lepton singlets.
Rotations in the space $(e_{jR},E_R)$ were already used to eliminate some terms.
Disregarding the three neutrinos masses, the Lagrangian \eqref{lag:E:gen} depends on $1(M_E)+9(Y^e)+5(Y^E)=15$ parameters, two more than the generic singlet vector-like \emph{quark} case\,\cite{nb-vlq} due to the Majorana phases in $Y^e=\pmnssm^\dag \hY^e$, $\pmnssm$ and $\hY^e$ being the PMNS matrix in the SM and the diagonal charged lepton Yukawa couplings.

We say the VLL is of Nelson-Barr type if the lagrangian \eqref{lag:E:gen} follows from a lagrangian of the form
\begin{equation}
\label{lag:E:NB}
\text{Singlet NB-VLL:\quad}
-\lag= \bar{\ell}_{iL} \cY^{e}_{ij} H e_{jR} + \bar{E}_{L} \cM^{Ee}_{rj} e_{jR} + \cM_E \bar{E}_{L} E_{R} +h.c.,
\end{equation}
where $\cY^\ell,\cM^E$ are real while $\cM^{Ee}$ is complex.
This Lagrangian depends on $1(\cM_E)+6(\cY^e)+5(\cM^{Ed})=12$ parameters which can be counted in the basis where $\cY^e=O_{e_L}\hat{\cY}^e$, $O_{e_L}$ and $\hat{\cY}^e$ being a real orthogonal matrix and a diagonal matrix of singular values, respectively.
So the entire CP violation in the lepton sector comes from $\cM^{Ee}$ which breaks CP softly as a consequence of spontaneous CP violation. 
Moreover, we will see that only one CP odd parameter will control CP violation so that correlations are expected between the Dirac CP phase, the two Majorana phases and other phases connecting the light-heavy mixing; see the quark case in \cite{nb-vlq}.
The structure of this breaking sector is unimportant here and it is assumed to be triggered at a much higher scale.
This definition can be based on the presence of a $\ZZ_2$ symmetry for which only $E_R,E_L$ are odd and 
only $\cM^{Ee}$ breaks this symmetry softly.\,\footnote{%
So spontaneous CP symmetry can be triggered by $\ZZ_2$ odd scalars.
}

For the case of NB-VLLs, the defining lagrangian \eqref{lag:E:NB} can be cast in the generic form \eqref{lag:E:gen}.
Hence, it is important to distinguish these two bases that will be used. We will refer to the basis in 
\eqref{lag:E:NB} as the \emph{real} basis in which the CP symmetry is manifest and it is only softly broken by $\cM^{eE}$.
In contrast, we will denote the form \eqref{lag:E:gen} as the \emph{generic} basis in which all VLLs, including the NB type, can be studied.

To be explicit about neutrino masses, in the singlet NB-VLL model in the real basis, we assume neutrino masses come from 
\eq{
\label{lag:nu}
-\lag\supset \frac{(\cM_{\nu})_{ij}}{v^2}(\ell_{iL}H)(\ell_{jL}H)\,,
}
where $\cM_\nu$ is real.
In general, we will use the basis where $\cM_\nu$ is diagonal.
In the generic basis, the structure is the same since $\ell$ do not change basis, and $M_\nu=\cM_\nu$.\,\footnote{%
In the case of a doublet NB-VLL with vector-like doublet $L_L,L_R$, this simple assumption is not valid because the basis change in the space $(\ell_L,L_L)$ will also modify neutrino masses. Thus only the singlet model will be considered here.
}

At last, let us make explicit how many parameters are CP odd.
In the SM with three massive Majorana neutrinos, there are already three CP violating phases: one Dirac CP phase and two Majorana phases.
When we add one generic VLL of charge $-1$, two more phases appear in clear analogy with the quark sector\,\cite{lavoura.branco,branco:book}.
In contrast, for the singlet NB-VLL case,
in total analogy with the case of one down-type vector-like quark of NB type\,\cite{nb-vlq}, there is \emph{only one} CP violating parameter that controls all CP violation.
We emphasize this information in table \ref{tab.1}.
From this counting, we expect the various CP violating phases to be highly correlated.
\begin{table}[h]
\[
\begin{array}{|c|cc|}
\hline
\text{SM+Majorana $\nu$}    & \text{\# of param.} & \text{\# of CP odd}\cr
\hline
\text{---}           & 9    & 3                \cr
+\ \text{1 gen-VLL}  & 15   & 5         \cr
+\ \text{1 NB-VLL}   & 12   & 1        \cr
\hline
\end{array}
\]
\caption{\label{tab.1}
Number of parameters in the lepton sector of the SM with Majorana neutrinos and with the addition of one VLL.
The light-neutrino masses are not counted.
}
\end{table}

%%%%%%%%%%%%%%%%%%%%%%%%%%%%%%%%%%%%%%%%%%%
\section{Parametrization}
\label{sec:diag.1}

To specify how NB-VLLs differ from generic VLLs, we need to change basis from the real basis in \eqref{lag:E:NB} to the generic basis in \eqref{lag:E:gen}.
This is easily described by comparing the $4\times 4$ mass matrix of the charged leptons following from \eqref{lag:E:NB} and \eqref{lag:E:gen}, respectively, after EWSB:
\eq{
\label{mass.matrix}
\text{real basis:}\quad\cM^{e+E}=\mtrx{\frac{v}{\sqrt{2}}\cY^e & 0\cr \cM^{Ee}  & \cM_E}
\,,\quad
\text{generic basis:}\quad M^{e+E}=\mtrx{\frac{v}{\sqrt{2}}Y^e & \frac{v}{\sqrt{2}}Y^E\cr 0 & M_E}\,.
}
Only a unitary transformation from the right, in the $3+1$ dimensional space $(e_{jR},E_R)$, is necessary to connect them.
% \eq{
% \cM^{e+E}W_R=M^{e+E}\,.
% }
The relevant exact relations are given by
\subeqali{
\label{ME}
M_E&=\sqrt{\cM^{Ee}{\cM^{Ee}}^\dag+\cM_E^2}\,,
\\
\label{YE:NB}
Y^E &=\cY^e w\,,
\\
\label{Ye:NB:S}
Y^e{Y^e}^\dag &=\cY^e\left(\id_3-ww^\dag\right){\cY^e}^\tp\,,
}
where we have defined the $3\times 1$ column vector
\eq{
w\equiv {\cM^{Ee}}^\dag/M_E\,.
}
One can show that $0<|w|<1$, cf.\,\cite{nb-vlq}.
The expression in eq.\,\eqref{Ye:NB:S} is the leading expression obtained analogously for the quark seesaw in BBP type models\,\cite{BBP,NB:CP4}.
It should be clear for the NB case that $Y^E$ in \eqref{YE:NB} and $Y^e$ in \eqref{Ye:NB:S} are not independent. For example, 
\eq{
\label{sum.rule}
	Y^e{Y^e}^\dag+Y^E{Y^E}^\dag=\cY^e{\cY^e}^\tp\sim \text{$3\times 3$ real},
}
and the imaginary part of $Y^E{Y^E}^\dag$ should coincide with the imaginary part of $-Y^e{Y^e}^\dag$.

The form $M^{e+E}$ is simpler to be analyzed due to the hierarchy $M_E\gg m_\tau$. In the leading seesaw approximation, we obtain the simple block diagonalization\,\footnote{%
In the (12) block we are neglecting $\ums{2}v^2Y^e{Y^e}^\dag\theta_L/M_E-\ums{2}\theta_L\theta_L^\dag\theta_LM_E$. The block (21) is one order smaller.
}
\eq{
\label{leading.ss}
U_{e_L}^\dag M^{e+E}\approx \mtrx{\frac{v}{\sqrt{2}}Y^e & 0\cr 0 & M_E}\,.
}
This means the blocks $vY^e/\sqrt{2}$ and $M_E$ in \eqref{mass.matrix} already give the light and heavy mass matrices at this order and they do not depend on $Y^E$ off the diagonal.
The latter only contributes at this order to the light-heavy mixing
\eq{
U_{e_L}\approx \mtrx{\id_3 & \theta_L\cr -\theta_L^\dag & 1}\,,
}
with
\eq{
\label{thetaL}
\theta_L\equiv \frac{v}{\sqrt{2}}Y^E/M_E\,.
}
This $3\times 1$ vector controls the seesaw expansion and should be small for the validity of the seesaw.

When the seesaw expansion is valid, $M_E$ is the VLL mass while $Y^e{Y^e}^\dag$ is the leading SM charged lepton Yukawa matrix squared which contains the PMNS mixing matrix:
\eq{
\label{Ye:sm.input}
Y^e{Y^e}^\dag =\frac{2}{v^2}V_{e_L}\diag(m^2_e,m^2_\mu,m^2_\tau)V_{e_L}^\dag\,.
}
The diagonalization matrix $V^\dag_{e_L}$ is the $3\times 3$ PMNS matrix $\pmnssm$, with Majorana phases from the right.
Note that the righthand side of \eqref{Ye:sm.input} contains 9 parameters from the SM lepton sector with Majorana neutrinos.
The three masses and three mixing angles are well known. The Dirac CP phase is planned to be measured in future experiments such as DUNE\,\cite{DUNE:2020ypp} while the two Majorana phases will be only indirectly constrained.
We will consider these 9 parameters as SM input, although the three CP phases are yet to be measured.

In the generic case, $Y^e$ and $Y^E$ are independent. Then, in the seesaw approximation, the relation \eqref{Ye:sm.input} completely fixes the 9 parameters in $Y^e$ as a function of the 9 parameters in the SM.
The Yukawa coupling $Y^E$ will contain 5 BSM parameters, two of which will be phases.

In the NB case, the lefthand side of \eqref{Ye:sm.input} needs to be generated from the special form \eqref{Ye:NB:S} and the same coupling $\cY^e$ also enters in $Y^E$ in \eqref{YE:NB}.
So correlations will appear and $Y^e$ and $Y^E$ are not independent\,\cite{nb-vlq}.
In special, the CP violation in the PMNS matrix should come entirely from \emph{one} CP odd parameter in $\cM^{Ee}$ (or $w$).

We now make these features explicit by describing the seesaw parametrization\,\cite{nb-vlq} first applied for the model with one down-type vector-like quark of Nelson-Barr type.

We first choose \eqref{ME} as one of the parameters instead of $\cM_E$ and rewrite
\eq{
\cM^{e+E}=\mtrx{\frac{v}{\sqrt{2}}\cY^e & 0\cr M_E w^\dag & M_E\sqrt{1-|w|^2}}\,.
}
We use the real rotation freedom on $e_{jR}$ to parametrize\,\cite{nb-vlq}
\eq{
\label{w}
w={\cM^{Ee}}^\dag/M_E=\mtrx{0\cr ib \cr a}\,,
}
with $b<a$ and $a^2+b^2< 1$.
So $\{M_E,w\}$ contains three parameters while $\cY^e$, being real and generic, contains 9 parameters.
This basis differs from the one used for parameter counting after eq.\,\eqref{lag:E:NB} but the total number of parameters is unchanged.
The CP properties are, however, more transparent in the current basis and only the $b$ parameter is CP odd.%
\footnote{For CP violation both $a,b$ need to be nonzero.}

Now, the combination in the righthand side of \eqref{Ye:NB:S} depends on $\cY^e$ and $w$, and should be partially fixed by SM input in \eqref{Ye:sm.input}.
The inversion formula is\,\cite{nb-vlq}
\eq{
\label{formula:cal-Yd}
\cY^e=A_1^{1/2}\cO \mtrx{1&&\cr &(1-b^2)^{-1/2}&\cr  &&(1-a^2)^{-1/2}}\,,
}
where $A_1$ is the real part in $Y^e{Y^e}^\dag=A_1+iA_2$.
The imaginary part enters in the calculation of the orthogonal matrix $\cO$ which is defined by
\eq{
\label{def:cO}
\cO^\tp A_1^{-1/2}A_2A_1^{-1/2}\cO
=\mu\mtrx{0&&\cr&0&-1\cr&1&0}\,,
}
where the righthand side is a canonical form for a real antisymmetric matrix and $0<\mu<1$.
The parameters $a$ and $b$ are not independent and are related by
\eq{
\label{def:mu}
\frac{a}{\sqrt{1-a^2}}\frac{b}{\sqrt{1-b^2}}=\mu\,.
}
The orthogonal matrix $\cO$ in \eqref{def:cO} is not unique as an additional rotation in the $23$ plane from the right will leave the equation invariant.
We fix a representative for $\cO$ using some convention and parametrize the additional freedom in the $23$ plane by an angle $\gamma$.

In this way, in our seesaw parametrization, among the 12 parameters in $\cM^{e+E}$, the 9 parameters, including the still unknown three CP phases, are fixed from the SM input while the following 3 BSM parameters are free: $\{M_E,b,\gamma\}$.
So $\cY^d$ and $w$ depend on $b,\gamma$, and $M_E$ can be chosen independently.
The angle $\gamma$ will be varied in the whole range $[0,2\pi]$. 
We can see from \eqref{def:mu} that $b$ cannot be arbitrarily small as $a/\sqrt{1-a^2}$ will be proportionally large.
We choose $b_{\min}\le b\le 1/\sqrt{2}$ with $b_{\min}=10^{-3}$ to set a lower limit.
For very small values of $b$, the seesaw approximation would not be precise when $\cY^e$ in \eqref{formula:cal-Yd}, or equivalently $Y^E$, is large.
See appendix \ref{ap:seesaw} for estimates of the deviations.
However, we will see in Sec.\,\ref{sec:params} that electroweak precision observables already constrain the Yukawa couplings $Y^E_i$ (or the mixing $V_{Ei}$) so that the seesaw approximation is good to within $0.2\%$.

We use the parametrization of the PMNS matrix, including Majorana phases $\beta_1,\beta_2$, in the form
\eq{
\label{pmns}
\pmnssm=V^\dag_{e_L}V_{\nu_L}=U(\theta_{12},\theta_{23},\theta_{13},\delta)\mtrx{1&&\cr &e^{i\beta_1}&\cr &&e^{i(\delta+\beta_2)}}\,.
}
The first piece is the standard parametrization. If we incorporate $e^{i\delta}$ of the second piece in the definition of this first piece, the first row is real.
We employ this parametrization for SM input in \eqref{Ye:sm.input}, in the basis where $V_{\nu_L}=\id_3$.
Similarly to the SM with Majorana neutrinos, we may consider $\beta_1,\beta_2\in [0,\pi)$ or $(-\pi/2,\pi/2]$ because sign flips $\ell_i\to -\ell_i$ in \eqref{lag:E:gen} can be used to absorb the additional signs from the left of $Y^e$ while $Y^E$ carries generic phases; neutrino masses in \eqref{lag:nu} are left invariant. 
For the NB-VLL in \eqref{lag:E:NB}, these sign flips can be used to set some sign conventions to $\cY^e$.

It is also useful to introduce 
\eq{
\label{def:Y-tilde}
\tY^E\equiv V_{e_L}^\dag Y^E=\pmnssm Y^E\,,
}
as the VLL Yukawa coupling in the basis where $Y^e$ is diagonal and the PMNS mixing of the SM accompanies the neutrinos.
Some phenomenological constraints are better expressed in terms of $\tY^E$.

\begin{figure}[h]
\includegraphics[scale=0.6]{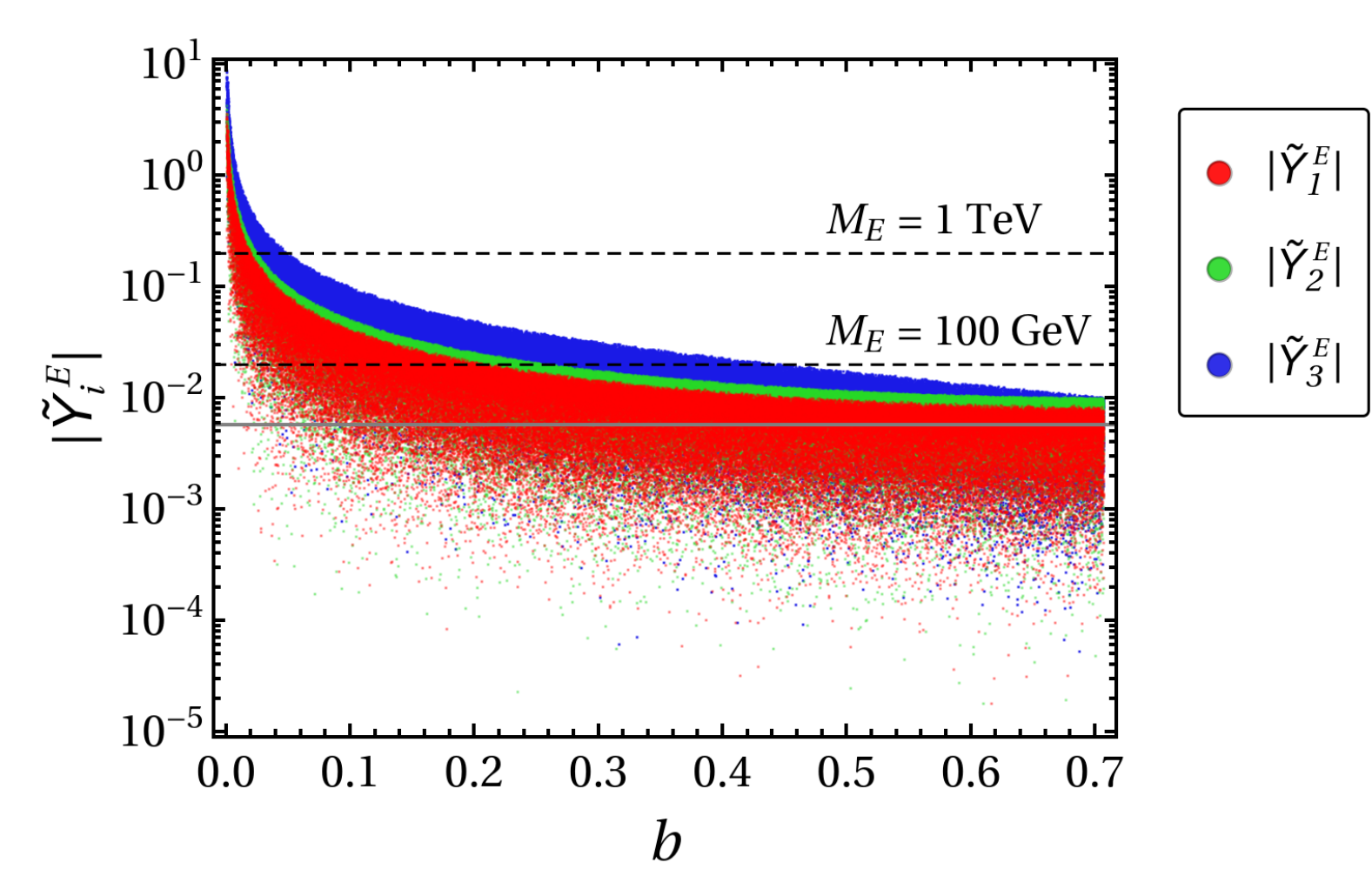}
\caption{\label{fig:Y-tilde}%
The Yukawa couplings $|\tY^E_i|$ as a function of $b$ in the seesaw parametrization.
The dashed lines mark electroweak precision constraints; cf.\,\eqref{tYE:bound}.
The continuous gray line is $y_\tau/\sqrt{3}$; cf.\,\eqref{tYE.ref}.
}
\end{figure}

To understand the flavor consequences of our seesaw parametrization that incorporates the SM input from \eqref{Ye:sm.input}, we show in Fig.\,\ref{fig:Y-tilde} the absolute values of $|\tY^E_i|$ as a function of $b$.
To produce this plot, we vary the leptonic mixing angles and Dirac CP phase in \eqref{pmns} within their 3$\sigma$ ranges resulting from the global fit in Ref.\,\cite{capozzi:global} while $b,\gamma$ are varied as discussed previously.
We can see that these components are all of the same order and no clear hierarchy is present.
This feature is clearly different from the case of a VLQ of Nelson-Barr type for which the Yukawa couplings inherit the hierarchy of the SM Yukawa couplings and mixing\,\cite{nb-vlq}. In the leptonic case, the non-hierarchical structure of $\tY^E_i$ is a consequence of the large mixing angles of the PMNS. The couplings $Y^E_i$ show similar non-hierarchical structure.

\begin{figure}[h]
\includegraphics[scale=0.45]{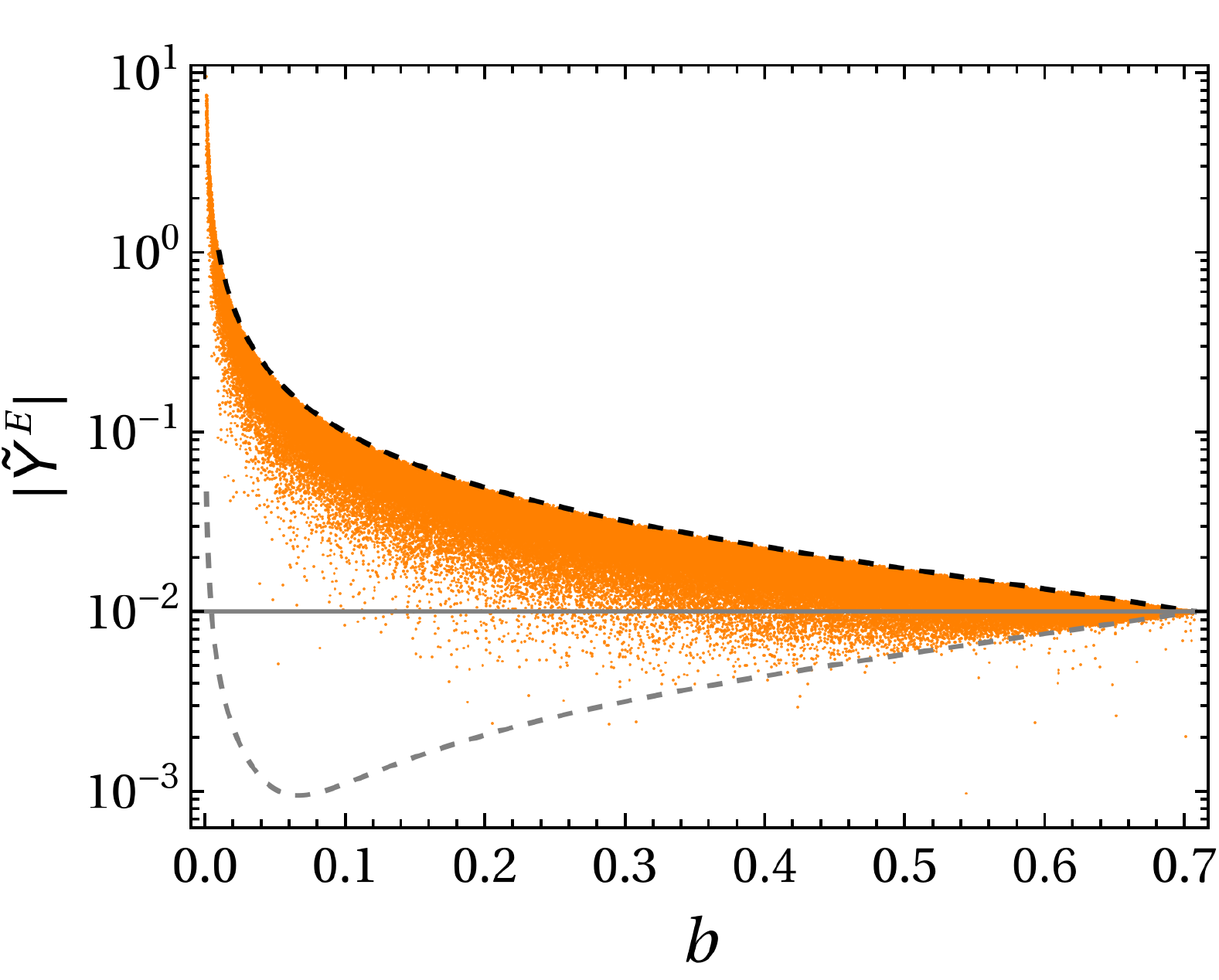}
\caption{\label{fig:Y-norm}
The norm of Yukawa couplings $|\tY^E|=\sqrt{|\tY^E_1|^2+|\tY^E_2|^2+|\tY^E_3|^2}=|Y^E|$ as a function of $b$ in the seesaw parametrization.
The black (gray) dashed curve marks the upper (lower) bound function in \eqref{YEmax:norm} [in  
\eqref{YEnorm:bounds} with $y_{\rm max}=y_\tau$ and $y_{\rm min}=4.5\times 10^{-5}$].
The continuous gray line is $y_\tau$; cf.\,\eqref{YE-norm.ref}.
}
\end{figure}

We also show in Fig.\,\ref{fig:Y-norm} the norm $|\tY^E|=|Y^E|$ as a function of $b$.
We can see that the points are approximately bounded from above by the black dashed curve and roughly bounded from below by the gray dashed curve. These curves are explained in appendix \ref{ap:seesaw}.

Comparing Fig.\,\ref{fig:Y-tilde} with Fig.\,\ref{fig:Y-norm}, we see that although $|Y^E_i|$ can take values over a wide range, the norm $|\tY^E|$ are roughly concentrated between the two dashed curves.
Therefore, when simple comparisons are necessary, we will adopt the reference value
\eq{
\label{YE-norm.ref}
|\tY^E|\sim y_\tau\sim 10^{-2}\,.
}
This value is shown in Fig.\,\ref{fig:Y-norm} as a continuous gray line.
Assuming equal values for all components $|\tY^E_i|$, the previous value translates into
\eq{
\label{tYE.ref}
\tY_0\equiv \frac{y_\tau}{\sqrt{3}}\sim 0.0058\,,
}
for $|\tY^E_i|$. This value is displayed in Fig.\,\ref{fig:Y-tilde} as a continuous gray line.

We should also stress that the property $|\tY^E_1|\sim |\tY^E_2|\sim |\tY^E_3|$ will not lead to couplings respecting minimal flavor violation in the lepton sector\,\cite{mfv:lep} if we integrate out the VLL.

%%%%%%%%%%%%%%%%%%%%%%%%%%%%%%%%%%%%%%%%%%%
\section{Observables: Couplings to $W,Z$ and higgs}
\label{sec:obs}

In the weak eigenstate basis, where the Yukawa Lagrangian in Eqs.\,\eqref{lag:E:gen} and \eqref{lag:E:NB} are valid, these couplings are given by
\eqali{
-\lag_{W}&=\frac{g}{\sqrt{2}}\bar{e}_{iL}\gamma^{\mu}\nu_{iL}W_{\mu}^{-}+h.c.
\cr
-\lag_{Z}&=\frac{g}{2c_W}\Big(\bar{\nu}_{iL}\gamma^{\mu} \nu_{iL}
-\bar{e}_{iL}\gamma^{\mu} e_{iL}-2s^2_W J^\mu_{e.m.}\Big)Z_\mu\,,
}
where the implicit sum over $i,j=1,2,3$ are for doublet leptons $\ell_{iL}=(\nu_{iL},e_{iL})^\tp$ and $J^\mu_{e.m.}$ is the electromagnetic current containing 
usual leptons and VLLs.
The coupling with the higgs can be read off from the Yukawa Lagrangian.

The couplings in the mass eigenstate basis are
\subeqali[lag:WZ:mass]{
\label{L:W:mass}
-\lag_{W}&=
\frac{g}{\sqrt{2}}\big(\bar{e}_{iL}V_{ij}+\bar{E}_{L}V_{4j}\big)\gamma^{\mu}\nu_{jL}
W_{\mu}^{-}+h.c.,
\\
\label{L:Z:mass}
-\lag_{Z}&=\frac{g}{2c_W}\left[\bar{\nu}_{iL}\gamma^{\mu}\nu_{iL}-\bar{e}_{iL}\gamma^{\mu}X_{ij}e_{jL}
-\bar{E}_{L}\gamma^{\mu}X_{44}E_{L}
\right.
\cr&\hspace{3em}
\left.
-2s_W^{2}J_{e.m.}^{\mu}
-\big(\bar{e}_{iL}\gamma^{\mu}X_{i4}E_{L}+h.c.\big)
\right]Z_{\mu}\,.
}
The $4\times 3$ PMNS mixing matrix in the charged current is 
\eq{
\label{V:singlet}
V=U^\dag_{e_L}P_eU_{\nu_L}
\approx
\left(\begin{array}{c}
\id_3-\ums{2}\delta X \cr\hline
\Theta^\dag
\end{array}\right)V_{e_L}^\dag\,,
}
where $P_e\sim 4\times 3$ is a projection matrix with $(P_e)_{ii}=1$ for $i=1,2,3$ and zero entries in the rest.
The last expression uses the seesaw approximation and assumes the basis where $U_{\nu_L}=\id_3$, with
\eq{
\label{Theta}
\Theta\equiv V_{e_L}^\dag \theta_L=\frac{v}{\sqrt{2}M_E}\tY^E\,,
\quad
\delta X\equiv \Theta\Theta^\dag\,,
}
where $V_{e_L}^\dag\approx \pmnssm$ is the $3\times 3$ PMNS matrix in the SM and 
$\theta_L$ is given in \eqref{thetaL}.

The square matrix $X$ of size $4$ that describes the FCNC coupling to $Z$ is
\eq{
\label{X^e}
X=V V^\dag
\approx 
\left(\begin{array}{c|c}
\id_3-\delta X & \Theta
\cr\hline
*  & \Theta^\dag\Theta
\end{array}\right)\,.
}
So we see that all couplings of VLLs to gauge bosons depend solely on the matrix $\Theta$ within the seesaw approximation.

Analogously, the higgs coupling in the mass eigenstate basis is
\eq{
-\lag_h=\frac{h}{v}\mtrx{\bar{e}_{iL}&\bar{E}_L}N^e\mtrx{e_{iR}\cr E_R}+h.c.\,,
\label{L:h:mass}
}
where
\eq{
N^e=X \hat{M}^{e+E}
\approx
\mtrx{\big(\id_3-\delta X\big)\hat{M}^e & 
	\Theta M_E
	\cr
	\Theta^\dag \hat{M}^e
	& 	\ums{2}\Theta^\dag\Theta M_E
}\,.
}
The hatted matrices are diagonalized matrices and the last expression is valid at leading order. 
The first term in the upper-left block is the standard coupling proportional to the lepton masses.

%%%%%%%%%%%%%%%%%%%%%%%%%%%%%%%%%%%%%%%%%%%
\section{Charged lepton flavor violation}
\label{sec:clv}

The mixing of the VLL with the SM charged leptons induces flavor changing interactions in the charged current with $W$ and in the neutral current with both the $Z$ in \eqref{lag:WZ:mass} and with the higgs in \eqref{L:h:mass}.
These flavor changing interactions will induce processes with charged lepton flavor violation (CLFV) which are tightly constrained.
We showed in Fig.\,\ref{fig:Y-tilde} that for one singlet NB-VLL of charge $-1$, its couplings to each of the SM charged leptons are typically of the same order of magnitude and large flavor changing effects may be possible.
We review the most relevant processes in the following.

\subsection{$\ell\to \ell'\ell''\ell''$}

Flavor changing decays $\ell_i\to \ell_j\ell_k\bar{\ell}_k$ or $\ell_i\to \ell_k\ell_k\bar{\ell}_k$ are induced at tree-level by the flavor changing current coupling to the $Z$ in \eqref{lag:WZ:mass}.
The latter involves the matrix $X_{ij}$. 
Following the analysis presented in Ref.\,\cite{Ishiwata:2013gma} we obtain the decay rates for the lepton $\ell_i$ decaying into three leptons:
\subeqali[eq:l.decay]{
\Gamma[\ell_i\rightarrow \ell_j\ell_k\bar{\ell}_k]&=\frac{G_{F}^{2}m_{i}^{5}|X_{ij}|^{2}}{768\pi^{3}}\left[(1-4 s_{w}^{2}+8 s_{w}^{4})+2(2 s_{w}^{2}-1)(1-X_{kk})+(1-X_{kk})^{2}\right]\;,
\\
\Gamma[\ell_i\to \ell_k\ell_k\bar{\ell}_k]&=\frac{G_{F}^{2}m_{i}^{5}|X_{ik}^{e}|^{2}}{384\pi^{3}}\left[(1-4 s_{w}^{2}+6 s_{w}^{4})+2(2 s_{w}^{2}-1)(1-X_{kk})+(1-X_{kk})^2\right]\;,
}
where $(\ell_1,\ell_2,\ell_3)=(e,\mu,\tau)$ and $i\neq j\neq k$ in the expressions. Of course, $\ell_i=\mu$ or $\tau$.
Note that $X_{ij}$ for $i\neq j$ and $1-X_{kk}$ would vanish if the VLL decouples.
We also consider the limit of massless product particles.
Apart from that, these equations are exact with respect to the mass diagonalization of the lepton masses. In the leading seesaw, which typically is a very good approximation, we would identify 
$X_{ij}=\delta_{ij}-\Theta_i\Theta_j^*$ in \eqref{Theta}.

As a last comment, the decay rates for $E\to \ell\ell'\bar{\ell'}$ and $E\to \ell q\bar{q}$ through $Z$ exchange are similar to \eqref{eq:l.decay} and go with $|X_{E\ell}|^2$, hence they typically do not show a preference for a specific flavor.

\subsection{ $\ell\to \ell'\gamma$}
\label{sec:muegamma}

Another relevant observable is the branching ratio $\text{Br}(\mu\rightarrow e \gamma)$ which is generated at one-loop level. Following an EFT analysis, for instance based on \cite{Crivellin:2018qmi,crivellin:global}, from the effective Hamiltonian
\eq{
\label{dipole}
\mathcal{H}_{eff}=c_{R}^{\ell_{f}\ell_{i}}\bar{\ell}_{f}\sigma_{\mu\nu}P_{R}\ell_{i}F^{\mu\nu} + \text{h.c},
}
one can derive the formulas
\eqali{
a_{\ell_{i}}=-4\frac{m_{\ell_{i}}}{e}\;\text{Re}\;c_{R}^{\ell_{i}\ell_{i}},\quad 
d_{\ell_{i}}=-2\;\text{Im}\;c_{R}^{\ell_{i}\ell_{i}}\\
\text{Br}(\ell_{f}\rightarrow \ell_{i} \gamma)=\frac{m_{\ell_{f}}^{3}}{4\pi\Gamma_{\ell_{f}}\hbar}\left(|c_{R}^{\ell_{i}\ell_{f}}|^{2}+|c_{R}^{\ell_{f} \ell_{i}}|^{2}\right).
}
We denote $a_{\ell_{i}}$ as the correction to the magnetic moment of the charged lepton $\ell_i$, while $d_{\ell_{i}}$ stands for its electric dipole moment. The coefficients $c_{R}^{\ell_{f}\ell_{i}}$ can be obtained directly by the generic Lagrangian connecting the new fermionic field with the SM vector gauge bosons and the higgs\;\cite{freitas,Stockinger:2006zn}
\eqali{
\lag_{V}&=\bar{\Psi}\left(\Gamma_{\Psi V}^{iL}\gamma_{\mu}P_{L}+\Gamma_{\Psi V}^{iR}\gamma_{\mu}P_{R}\right)\ell_{i}V^{*}_{\mu}+\text{h.c.}\\
\lag_{h}&=\bar{\Psi}\left(\Gamma_{\Psi h}^{iL}P_{L}+\Gamma_{\Psi h}^{iR}P_{R}\right)\ell_{i}h+\text{h.c.}\\
}
In our model only the neutral left-current is modified by the presence of the singlet VLL ($\Gamma_{\Psi V}^{iR}=0$), which implies that the contribution to $c_{R}^{\ell_{f}\ell_{i}}$ induced by the VLL coupling to the $Z$ boson is
\eq{
c_{ZR}^{\ell_{f}\ell_{i}}=\frac{e}{16\pi^{2}}\left(m_{i}\Gamma_{E Z}^{fL*}\Gamma_{E Z}^{iL}\right)\frac{\tF_{V}\left(\frac{M_{E}^{2}}{m_{Z}^{2}}\right)}{m_{Z}^{2}}\,.
}    
The value for $\Gamma_{E Z}^{iL}$ can be read from \eqref{L:Z:mass} while $\tF_V(x)\equiv\tf_V(x)-\tg_V(x)$
is the combination of the loop functions\,\cite{Crivellin:2018qmi}:
\eqali{
\tilde{f}_{V}(x)&=\frac{-4x^{4}+49x^{3}-78x^{2}+43x-10-18x^{3}\log x}{24(x-1)^{4}}\nonumber\,,\\
\tilde{g}_{V}(x)&=\frac{-3(x^{3}-6x^{2}+7x-2+2x^{2}\log x)}{8(x-1)^{3}}\,.
}
The contribution due to the VLL coupling to the higgs boson can be obtained in a similar fashion. Notice that the couplings with both chiralities appear
\eq{
c_{hR}^{\ell_{f}\ell_{i}}=\frac{e}{16\pi^{2}}\left(m_{i}\Gamma_{E h}^{fL*}\Gamma_{E h}^{iL}\right)\frac{\tF_{S}\left(\frac{M_{E}^{2}}{m_{h}^{2}}\right)}{m_{h}^{2}}+\frac{e}{16\pi^{2}}\left(M_{E}\Gamma_{E h}^{fL*}\Gamma_{E h}^{iR}\right)\frac{F_{S}\left(\frac{M_{E}^{2}}{m_{h}^{2}}\right)}{m_{h}^{2}}\,.
}
From \eqref{L:h:mass}, it is easy to see that $\Gamma_{E h}^{iL}\propto M_{E}$, and $\Gamma_{E h}^{iR}\propto m_{i}$, 
making the two contributions above to be of the same order (the prefactors are in fact the same).
This property also justifies the absence of the term $\Gamma_{E h}^{fR*}\Gamma_{E h}^{iR}$. 
The functions $\tF_S(x)\equiv \tf_S(x)-\tg_S(x)$ and $F_S(x)\equiv f_S(x)-g_S(x)$ are the combination of the loop functions\,\cite{Crivellin:2018qmi}:
\eqali{
f_{S}(x)&=\frac{x^{2}-1-2x\log x}{4(x-1)^{3}}\,,\quad
g_{S}(x)=\frac{x-1-\log x}{2(x-1)^2}\,,\nonumber\\
\tilde{f}_{S}(x)&=\frac{2x^{3}+3x^{2}-6x+1-6x^{2}\log x}{24(x-1)^{4}}
\,,\quad
\tilde{g}_{S}(x)=\frac{1}{2}f_S(x)
\,.
}

In our model, the presence of the VLL will also modify the couplings of SM particles among themselves, implying that the process $\ell_{i}\rightarrow \ell_{f}+\gamma$ will be induced by diagrams containing only SM fields as well.\footnote{%
The process is similar to the negligible radiative neutrino decay in the SM with massive neutrinos.
Polarization asymmetry of photons, however, may result in the presence of new CP violating couplings and mediators\,\cite{Balaji:2019fxd}.
} 
By taking into account all contributions, including the latter, we can obtain the coefficient $c_R^{\ell_f\ell_i}$  of the dipole operator which we show in appendix \ref{ap:formulas}.
Using this coefficient, we obtain the final formulas for the observables
\eqali{
\label{Br:mu.e.gamma}
\text{Br}(\ell_{i}\rightarrow \ell_{f} \gamma)&=\frac{2\alpha}{(4\pi)^{4}}\frac{m_{\ell_{i}}^{5}}{\Gamma_{\ell_{i}}\hbar}G_{F}^{2}
\left(1+\frac{m_{\ell_{f}}^{2}}{m_{\ell_{i}}^{2}}\right)\left|X_{fi}\right|^{2}
\left[\tF_{V}\left(x_{E}\right)
\right.
\cr
&\hspace{10em}\left.
+\ y_{E}\left(\tF_{S}\left(y_{E}\right)+F_{S}\left(y_{E}\right)\right)
+\ \frac{1}{6}-\frac{2}{3}s_{w}^{2}\right]^{2}
\,,
\\
\delta a_{\ell_i}&=-4\sqrt{2}\frac{m_{\ell_i}^{2}}{(4\pi)^{2}}G_{F}\left|X_{Ei}\right|^{2}
\left[\tF_{V}\left(x_{E}\right)+y_{E}\left(\tF_{S}\left(y_{E}\right)+F_{S}\left(y_{E}\right)\right)+\frac{1}{6}-\frac{2}{3}s_{w}^{2}\right]
\,,
}
where $x_{E}\equiv\frac{M_{E}^{2}}{m_{Z}^{2}}$ and $ y_{E}\equiv \frac{M_{E}^{2}}{m_{h}^{2}}$.
We considered the seesaw approximation 
$X_{4f}^*X_{4i}\simeq \delta_{fi}-X_{fi}$
when writing the final formulas.
See appendix \ref{ap:formulas} for the coefficient $c_R^{\ell_f\ell_i}$ without this approximation.
The formula for $\text{Br}(\ell_{i}\rightarrow \ell_{f} \gamma)$ matches Ref.\,\cite{crivellin:global}, except for the contribution proportional to $F_S(y_E)$ and the one proportional to $s^2_w$.
These come from diagrams involving $\bar{E}_L\mu_Rh$ and $\bar{\mu}_RZ\mu_R$, respectively.
See appendix \ref{ap:formulas} for more details. 
This difference makes the constraint on radiative decays for $M_E=100\,\unit{GeV}$ 6.3 times weaker compared to Ref.\,\cite{crivellin:global}.
The formula for $\delta a_{\ell_i}$, which is very similar, matches Ref.\,\cite{freitas}.
We note that for fixed $X_{fi}$, the dependence of these expressions on the mass $M_E$ is not strong.

We have presented the expression for $\delta a_{\ell_i}$ just for completeness: 
although $\delta a_\mu$ can be positive, it cannot be large enough to explain the long standing deviation\,\cite{Bennett:2006fi} which was recently confirmed\,\cite{Muong-2:2021ojo}.
For example, for $M_E=500\,\unit{GeV}$, we would need the large and unphysical value $|X_{E\mu}|^2\sim 11$ to explain $a_\mu$.
In fact, a single VLL in any representation that mixes with the SM cannot explain $a_\mu$\,\cite{freitas}.
With two representations, there are many working options\,\cite{freitas,Crivellin:2018qmi,strumia}.
Since $c_{R}^{\ell_{i}\ell_{i}}$ is real, the VLL also does not contribute to the electric dipole moment at one-loop level.\footnote{%
It may contribute at two loops\,\cite{Fabbrichesi:1987sy}.
}

As a last remark, we can compare the expression for the radiative decays $\ell_i\to\ell_f\gamma$ in \eqref{Br:mu.e.gamma} to the prediction of minimal flavor violation in the lepton sector\,\cite{mfv:lep}.
Considering that for the NB-VLL typically $|X_{e\mu}|\sim |X_{e\tau}|\sim |X_{\mu\tau}|$, the decay rates for the NB-VLL will not follow the MFV prediction: $\Gamma(\mu\to e\gamma)/\Gamma(\tau\to \mu\gamma)\sim 0.03\, m_\mu^5/m_\tau^5$ and $\Gamma(\mu\to e\gamma)/\Gamma(\tau\to e\gamma)\sim m_\mu^5/m_\tau^5$.
For the NB-VLL, the factor 0.03 in the first ratio is also unity.

\subsection{Conversion in nuclei}

The flavor changing current coupling to the $Z$ in \eqref{L:Z:mass} also induces the flavor conversion 
$\mu^-N\to e^-N^{(*)}$ for a muon bound to a nucleous $N$.
Giving the tree-level nature of the latter, the limits obtained from this kind of process are expected to outperform the limits from all the other charged lepton violating process in the $\mu e$ sector\,\cite{gouvea.vogel}.
The best current limit comes from conversion in gold nuclei, where the experimental bound, when normalized by the capture rate for $\mu^-N\to\nu_\mu N'$, gives us $\Gammaconv/\Gammacapt{Au} < 7.0 \times 10^{-13}$, where $\Gammacapt{Au}= 8.7 \times 10^{-18}$ GeV \cite{SINDRUMII:2006dvw}.  
The expression for the conversion inside the nucleous $N$ can be written as\,\cite{crivellin:global}:
\begin{equation}
	\Gammaconv = 4 m_\mu^5 \left| \sum_{q=u,d} \left( C^{V\,LR}_{qq} + C^{V\,LL}_{qq} \right) \left( f^{(q)}_{V_p} V^p_N + f^{(q)}_{V_n} V^n_N \right) \right|^2,
	\label{eq:nucleiconversion}
\end{equation}
where $f^{(u)}_{V_p} = 2$, $f^{(u)}_{V_n} = 1$, $f^{(d)}_{V_p} = 1$ and $f^{(d)}_{V_n} = 2$ are the nucleon vector form factors. 
The $C$ coefficients are Wilson coefficients of effective four-fermion operators such as $(\bar{e}\gamma^\mu P_L\mu)(\bar{q}\gamma_\mu P_Lq)$.
In the VLL model, they are given by
\begin{equation}
	C^{V\,LR}_{qq} = \frac{g_W}{2 c_W} \frac{X_{e\mu}}{M_Z^2} \Gamma^R_{qq}, \qquad	C^{V\,LL}_{qq} = \frac{g_W}{2 c_W} \frac{X_{e\mu}}{M_Z^2} \Gamma^L_{qq},
\end{equation}
\begin{equation}
	\Gamma^L_{uu} = - \frac{g_W}{c_W} \left( \frac{1}{2} - \frac{2}{3} s_W^2\right), \quad \Gamma^R_{uu} =  \frac{2}{3}  \frac{g_W s_W^2}{c_W},
\end{equation}
\begin{equation}
	\Gamma^L_{dd} = - \frac{g_W}{c_W} \left( -\frac{1}{2} + \frac{1}{3} s_W^2\right), \quad \Gamma^R_{dd} =  -\frac{1}{3}  \frac{g_W s_W^2}{c_W}\,.
\end{equation}

\noindent
Using, for the gold nucleous, $V^p_{\rm Au} = 0.0974$ and $V^n_{\rm Au} = 0.146$, we find
\begin{equation}
	\Gammaconv = 2.8153 \times 10^{-16} |X_{\mu e}|^2 \,\unit{GeV}\,.
\end{equation}
The above expression, combined with the experimental bound, leads to the upper limit:
\begin{equation}
    |X_{\mu e}| < 1.5 \times 10^{-7},
\end{equation}
giving us the strongest constraint on this quantity.

The future Mu2e\,\cite{mu2e} and COMET\,\cite{comet} experiments are expected to dramatically improve the constraint and reach $\Gammaconv/\Gammacapt{Al} < 3 \times 10^{-17}$ using aluminium nuclei.
Given the different nucleus, we can use the estimate shown in Refs.\,\cite{ligeti.wise,gouvea.vogel} which yields
\eq{
\label{mu2e.limit}
|X_{\mu e}|< 4\times 10^{-9}\,.
}

\subsection{Current limits}
\label{sec:current.lim}

We collect in table \ref{tab:LFV} the limits coming from lepton flavor violating processes discussed in this section. We also show the corresponding limit on $X_{ij}$, $i\neq j$, within the seesaw approximation for which
\eq{
X_{i4}X_{j4}^*\simeq -X_{ij}\simeq\Theta_i\Theta_j^* = \frac{v^2}{2M_E^2}\tY^E_i\tY^{E*}_j
\,,\quad
i,j=1,2,3.
}
The Yukawa couplings $\tY^E_i$ were defined in \eqref{def:Y-tilde}.
In the same table we also show the future limits for the same observables. 
The constraints for $|X_{ij}|$ can be obtained by rescaling by the square root of the ratio between the future and current limits on the branching ratios.

\begin{table}[h]
\[
\begin{array}{|c|c|c|c|}
\hline  \text{Observable}    & \text{Current limit} & \text{Constraint} & \text{Future limit} 
    \cr
\hline
\text{Br}(\mu\to eee)       &      <1.0 \times 10^{-12} \text{\cite{PDG2020}}       &   
    |X_{\mu e}|< 2.2 \times 10^{-6} & 10^{-16} \text{\cite{mu3e}} \cr
% \text{Br}(\tau\to \mu\mu\mu)       &    2.1 \times 10^{-8} \text{\cite{PDG2020}}?            &         \cr
\text{Br}(\tau\to \mu\mu\mu)       &   <2.1 \times 10^{-8} \text{\cite{HFLAV:2019otj}}            &
    |X_{\tau\mu}|< 5.6 \times 10^{-4} & 4\times 10^{-10} \text{\cite{belle2:book}} \cr
\text{Br}(\tau\to \mu ee)       &  <8.4 \times 10^{-9} \text{\cite{HFLAV:2019otj}}              &
    |X_{\tau\mu}|< 6.2 \times 10^{-4} & 3\times 10^{-10} \text{\cite{belle2:book}}\cr    
% \text{Br}(\tau\to eee)       &  2.7 \times 10^{-8} \text{\cite{PDG2020}?              &        \cr
\text{Br}(\tau\to eee)       &  <1.4 \times 10^{-8} \text{\cite{HFLAV:2019otj}}            &
    |X_{\tau e}|< 6.3 \times 10^{-4} & 5\times 10^{-10} \text{\cite{belle2:book}}\cr
\text{Br}(\tau\to e\mu\mu)       &  <1.6 \times 10^{-8} \text{\cite{HFLAV:2019otj}}             &
    |X_{\tau e}|< 8.5 \times 10^{-4} & 5\times 10^{-10} \text{\cite{belle2:book}}\cr
\hline
\text{Br}(\mu\to e\gamma)       &  <4.2 \times 10^{-13} \text{\cite{PDG2020}}            & 
    |X_{\mu e}| < 9.5\times 10^{-5}& 6\times 10^{-14}\text{\cite{meg2}} \cr
\text{Br}(\tau\to \mu\gamma)       &  <4.4 \times 10^{-8} \text{\cite{PDG2020}}            &
    |X_{\tau\mu}| < 7.3 \times 10^{-2} & 10^{-9} \text{\cite{belle2:book}} \cr
\text{Br}(\tau\to e\gamma)       &  <3.3 \times 10^{-8} \text{\cite{PDG2020}}          &  
    |X_{\tau e}| < 6.3 \times 10^{-2}& 3\times 10^{-9} \text{\cite{belle2:book}} \cr 
\hline
\Gamma^{\rm conv}_{\mu \rightarrow e}/\Gamma^{\rm capt}_{\rm Au} & <7.0 \times 10^{-13} \text{\cite{SINDRUMII:2006dvw}} &
    |X_{\mu e }| < 1.5 \times 10^{-7} & 3 \times 10^{-17} \text{\cite{mu2e}}^{*}\cr
\hline
\end{array}
\]
\caption{\label{tab:LFV}
Current and future limits for charged lepton flavor violating processes at 90\% CL.
The constraints on radiative decays assume $M_E=100\,\unit{GeV}$.
To extract the future constraints for $|X_{ij}|$, rescale by the square root of the ratio between the future and current limits on the branching ratios.
*\,The nucleous is Al and rescaling is not valid; cf.\,\eqref{mu2e.limit}.
}
\end{table}

For the tree level decays, the sole dependence is on the mixing $X_{ij}$.
For radiative decays, there is an additional weak dependence on the VLL mass; cf.\,\eqref{Br:mu.e.gamma}.
Therefore, we show the limits for $M_E=100\,\unit{GeV}$. For $M_E=1\,\unit{TeV}$, these constraints
on $|X_{ij}|$ are weaker and should be multiplied by 1.8. Between these values, the limits are even weaker because the branching ratio vanishes for $M_E\sim 239\,\unit{GeV}$. See function in appendix \ref{ap:formulas}.

We can see that for $|\tY^E_1|\sim |\tY^E_2|\sim |\tY^E_3|$, the strongest constraint comes from $\mu\to e$ conversion in gold.
In terms of the Yukawa couplings $\tY^E_e$ and $\tY^E_\mu$, we find
\eq{
\label{constraint:Yemu}
|\tY^E_\mu\tY^{E*}_e|< 5.0\times 10^{-8}\left(\frac{M_E}{100\,\unit{GeV}}\right)^2\,,
}
or, relative to the reference value \eqref{tYE.ref}, 
\eq{
\frac{\sqrt{|\tY^E_\mu\tY^{E*}_e|}}{\tY_0}< 0.039\times \frac{M_E}{100\,\unit{GeV}}\,.
}
Therefore, for $M_E=100\,\unit{GeV}$, the couplings in Fig.\,\ref{fig:Y-tilde} are severely constrained.
As the mass increases, the constraint is weakened linearly and the gray line is reached only for 2.6\,\unit{TeV}. So typically masses below this value are excluded unless the couplings $\tY^E_\mu$ or $\tY^{E}_e$ are highly suppressed.

The constraints for flavor conversion in other sectors are much weaker. The no observation of the processes $\tau\to\mu\mu\mu$ and $\tau\to eee$ lead to
\eqali{
\frac{\sqrt{|\tY^E_\tau\tY^{E*}_\mu|}}{\tY_0}&< 2.4\times \frac{M_E}{100\,\unit{GeV}}\,,
\cr
\frac{\sqrt{|\tY^E_\tau\tY^{E*}_e|}}{\tY_0}&< 1.4\times \frac{M_E}{100\,\unit{GeV}}\,.
}
The gray line in Fig.\,\ref{fig:Y-tilde} comfortably passes these constraints for $M_E>100\,\unit{GeV}$.

Let us briefly comment on other CLFV processes.
The limits from flavor changing decays $h\to \ell\ell'$ and $Z\to \ell\ell'$ are currently weaker or at most similar to the constraints discussed above.
The flavor changing $Z$ decays for example currently constrain $|X_{\ell\ell'}|\lesssim 10^{-3}$.
In the future Tera $Z$ factories are expected to reach a sensitivity similar of low energy CLFV observables\,\cite{calibbi.marcano} in the channels $Z\to \tau\ell$.
Lepton flavor universality violation tests are already considered in the fit of Ref.\,\cite{crivellin:global} and leads currently to similar constraints such as $||\Theta_1|^2-|\Theta_2|^2|\lesssim 10^{-3}$, with similar values for the $\tau$ flavor\,\cite{ligeti.wise}.

%%%%%%%%%%%%%%%%%%%%%%%%%%%%%%%%%%%%%%%%%%%
\section{Special points of parameters}
\label{sec:special}

\subsection{Full decoupling of NB-VLL and CP violation}

Differently from the quark sector, CP violation in the leptonic sector has not been established yet, despite strong preference towards CP violating values for $\delta$ being reported by the T2K experiment\,\cite{t2k.2020}. The global fit of Ref.\,\cite{capozzi:global} still allows, for normal ordering, $\delta=\pi$ within $2\sigma$ and $\delta=0$ within $3\sigma$.

This unlikely scenario of CP conservation in the leptonic sector corresponds to taking all couplings in the Lagrangian \eqref{lag:E:NB} to be real.\,\footnote{%
This implies trivial Majorana phases as well.
}
Change of basis of the real basis to the generic basis \eqref{mass.matrix} still maintains the couplings real and we can decouple the NB-VLL from the SM by simply turning off $Y^E=0$ (or $\cM^{Ee}=0$).
In the decoupled SM, the PMNS matrix would be real.

Once leptonic CP violation is established, the NB-VLL can no longer decouple from the SM. The reason is that CP violation, which is (softly) broken only by $\cM^{Ee}$ in \eqref{lag:E:NB}, needs to be transmitted to the SM. This feature can be clearly seen in \eqref{Ye:NB:S}, which we rewrite as
\eq{
\label{NB:Ye:equality}
\cY^e\left(\id_3-ww^\dag\right){\cY^e}^\tp
% = Y^e{Y^e}^\dag 
=V_{e_L}\diag(y^2_e,y^2_\mu,y^2_\tau)V_{e_L}^\dag
\,,
}
using \eqref{Ye:sm.input}.
The righthand side is $Y^e{Y^e}^\dag$ and depends on $V^\dag_{e_L}=\pmnssm$, the $3\times 3$ PMNS matrix of the SM, which 
contains the CP violation in the SM. This CP violation needs to be generated by $w={\cM^{Ee}}^\dag/M_E$ of order one in the lefthand side. So $\cM^{Ee}$ cannot be suppressed compared to $M_E$.
This is analogous to what happens when quark CP violation is transmitted by vector-like quarks of NB type\,\cite{nb-vlq,vecchi.1}.

One can quantify the decoupling of the NB-VLL with\,\cite{vecchi.1}
\eq{
\label{def:mixbare}
\frac{1}{\cM_E}{\cM^{Ee}}^\dag=(Y^e)^{-1}Y^E=\frac{w}{1-|w|^2}\,.
}
This is the ratio between the mixing term of the NB-VLL with the SM and its bare mass, and it vanishes if the NB-VLL decouples, i.e., for $\cM^{Ee}=0$. 
Taking the norm, in the last form, this quantity is minimized when $|w|$ is maximized.
The norm $|w|=\sqrt{a^2+b^2}$ only depends on $b$ and $\mu$ through \eqref{def:mu}. 
A simple calculation shows that $|w|$ is maximal when $b=a=\sqrt{\frac{\mu}{1+\mu}}$, i.e., $\mu$ should be maximal as well.
In turn, $\mu$ is only a function of the Yukawa couplings in \eqref{def:cO}, including the Majorana phases $\beta_i$.

\begin{figure}[h]
\includegraphics[scale=0.45]{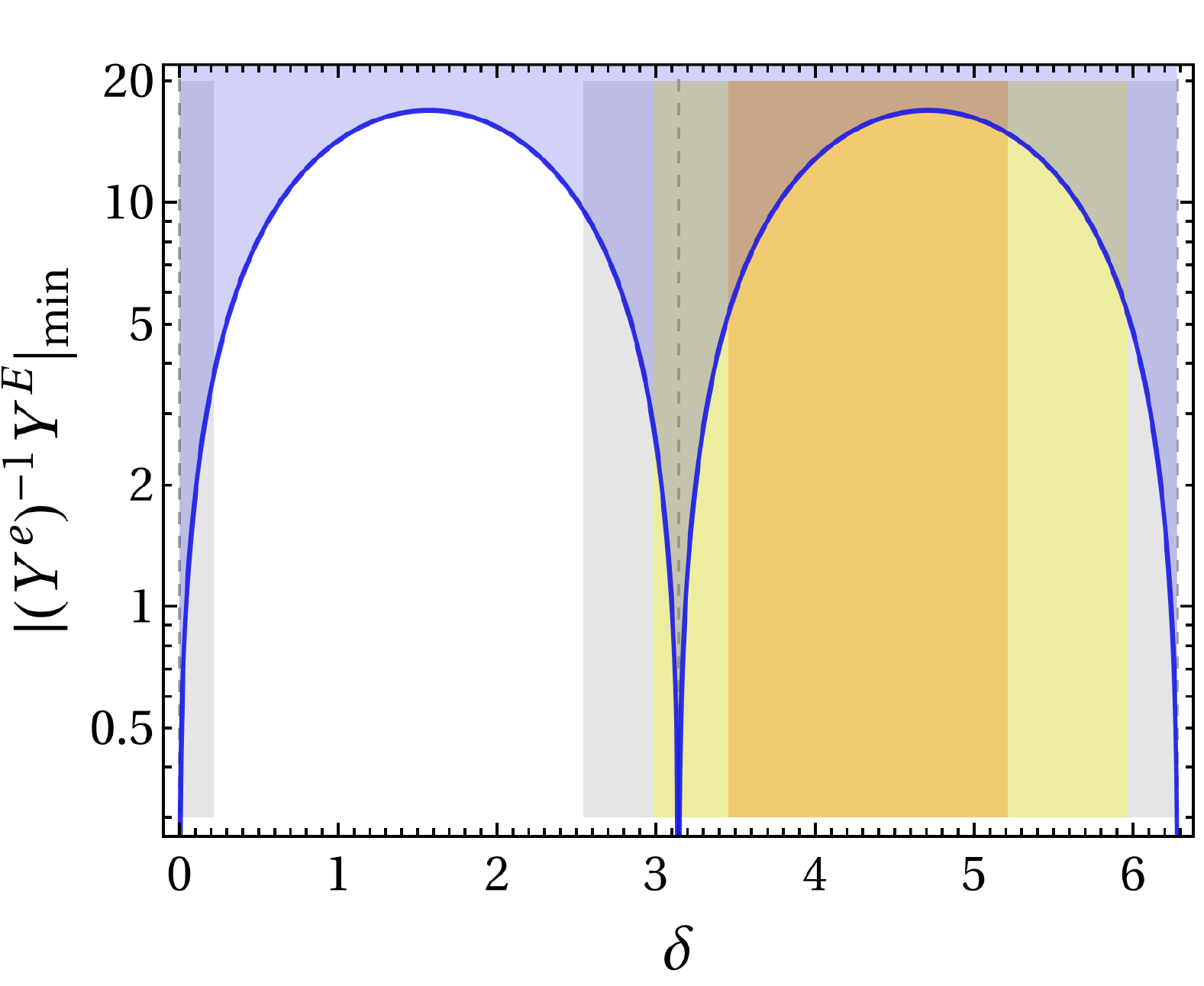}
\caption{\label{fig:mixbare}%
Quantifier of decoupling, norm of \eqref{def:mixbare}, as function of $\delta$.
The blue curve is the minimal value, minimized with respect to $b$ and the Majorana phases $\beta_i$.
The mixing angles are fixed to the best-fit values of Ref.\,\cite{capozzi:global}.
The gray, yellow and orange bands show the $3\sigma$, $2\sigma$ and $1\sigma$ bands for $\delta$, respectively, from the same fit.
}
\end{figure}

We show in Fig.\,\ref{fig:mixbare} the norm of the quantity in \eqref{def:mixbare}, minimized with respect to $b$ and the Majorana phases, as a function of the Dirac CP phase $\delta$. The blue curve depicts the minimal values and values above the curve are allowed. We clearly see that decoupling is only possible for the CP conserving values $\delta=0,\pi$. 
Away from these points, one quickly needs a large mixing term to reproduce the leptonic CP violation in the SM.
For example, if we are restricted to the $1\sigma$ band (orange) for $\delta$\,\cite{capozzi:global}, we need the ratio \eqref{def:mixbare} to have a norm larger than 5.
Since the CP violating mixing term and the CP conserving bare mass are in principle disconnected in origin, this raises the issue of a coincidence of scales that can be justified in a more complete construction\,\cite{vecchi.2}.
For comparison, this minimal value is 2 and 20 for a vector-like quark of Nelson-Barr type of down or up type, respectively\,\cite{vecchi.1}.

\subsection{NB-VLL decoupling of one flavor}
\label{sec:1flavor=0}

Although the NB-VLL cannot decouple from the SM in the presence of leptonic CP violation, we will show here that quite generically the NB-VLL can \emph{always} avoid coupling to \emph{one} of the lepton flavors for special values of the parameters $\beta_1,\beta_2$ in \eqref{pmns}.
As a consequence, the flavor changing neutral current coupling to the $Z$ boson does not involve that decoupled flavor at tree-level.
The situation can be summarized through the patterns
\eq{
\label{tYE:special}
(\tY^E_e,\tY^E_\mu,\tY^E_\tau)=(0,\times,\times)
\text{~~or~~}
(\times,0,\times)
\text{~~or~~}
(\times,\times,0)
\,.
}
Conversely, these special values for the betas are the only way to have one vanishing $\tY^E_i$ independently of the parameter $b$.
So if the PMNS mixing accompanies the neutrinos, the heavy lepton $E$ does not couple to one of the lepton flavors.
Note that these choices for $\beta_1,\beta_2$ are special and typically, as shown in Fig.\,\ref{fig:Y-tilde}, the NB-VLL couples to all flavors with couplings of the same order.
We also note these special situations are also present for one singlet vector-like quark of NB type of both up or down type.

To show the first part, let us first define the columns
\eq{
\label{VeL:ui}
V_{e_L}=\pmnssm^\dag =
    \left(
    \begin{array}{c|c|c}
    u_1 & u_2 & u_3
    \end{array}
    \right)\,,
}
corresponding to the complex conjugate of \emph{rows} of the PMNS matrix.
The vectors $u_i$ are the eigenvectors of $Y^e{Y^e}^\dag$:
\eq{
\label{YYe:ui}
Y^e{Y^e}^\dag u_i=y_i^2u_i\,,
}
for $i=1,2,3$ or $e,\mu,\tau$.

We note next that varying the Majorana phases $\beta_1,\beta_2$ in \eqref{pmns} is equivalent to \emph{rephasing} $Y^e{Y^e}^\dag$ and the components of $u_i$. So it is always possible to choose one of the $u_i$ to be real by adjusting the betas and overall rephasing. 
Let us choose $u_3$ to be real in the following. 
This corresponds to having the third row of the PMNS matrix real. 
A different choice works similarly.
For complex PMNS with nontrivial $\delta$, the eigenvectors $u_1,u_2$ should be complex because if one of them is also real, then the other one will be equally real by orthogonality and the PMNS would be real.

If the eigenvector $u_3$ is real, its eigenvector equation \eqref{YYe:ui} can be decomposed as
\eq{
\label{real.u3}
\re(Y^e{Y^e}^\dag) u_3=y_3^2u_3\,,\quad
\im(Y^e{Y^e}^\dag) u_3=0\,.
}
In the notation of \eqref{formula:cal-Yd}, $u_3$ is also an eigenvector of $A_1^{-1/2}A_2A_1^{-1/2}$ with eigenvalue zero.
This means $\cO$ in \eqref{def:cO} has $\pm u_3$ at its first column and then it follows from \eqref{formula:cal-Yd} that
\eq{
\label{u3.cYe}
u_3^\dag \cY^e=\pm y_3\,(1,0,0)\,.
}
From \eqref{YE:NB} and \eqref{w},
\eq{
\tY^E_3=(V_{e_L}^\dag Y^E)_3=u_3^\dag Y^E=u_3^\dag \cY^ew=0\,.
}
This is the third pattern in \eqref{tYE:special}.
If we had chosen another row of the PMNS matrix to be real, the corresponding $\tY^E_i$ would vanish.

Additionally, it follows from \eqref{sum.rule} that
\eq{
\cY^e{\cY^e}^\tp u_3=(Y^e{Y^e}^\dag+Y^E{Y^E}^\dag)u_3=y_3^2 u_3\,,
}
i.e., $u_3$ is also an eigenvector of $\cY^e{\cY^e}^\tp$ with eigenvalue $y_3^2$.
Combining with \eqref{u3.cYe}, we can conclude that the first column of $\cY^e$ is $\pm y_3u_3$ and the other columns are orthogonal to $u_3$.

\begin{figure}[h!]
\includegraphics[scale=0.55]{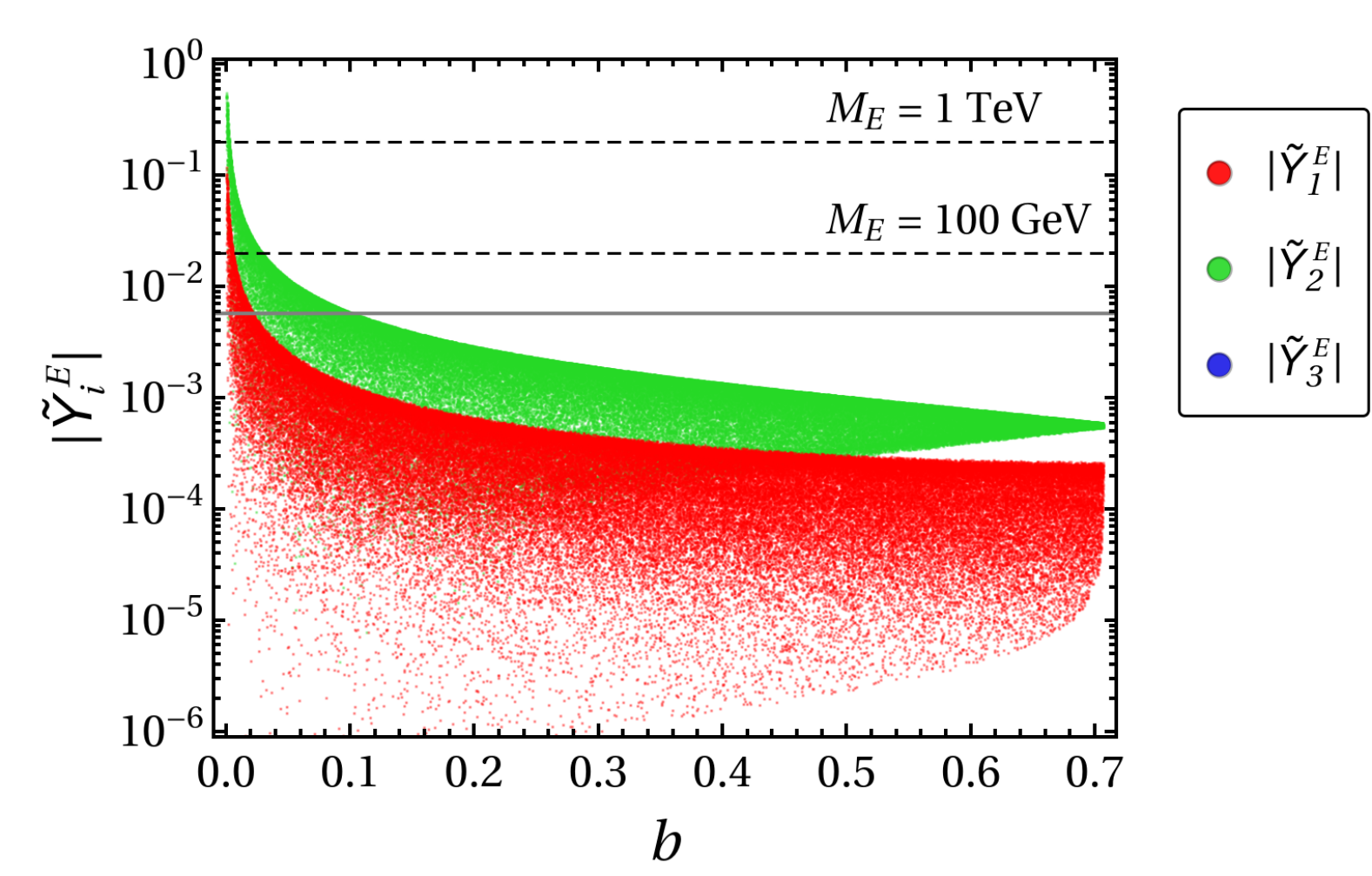}
~~
\includegraphics[scale=0.44]{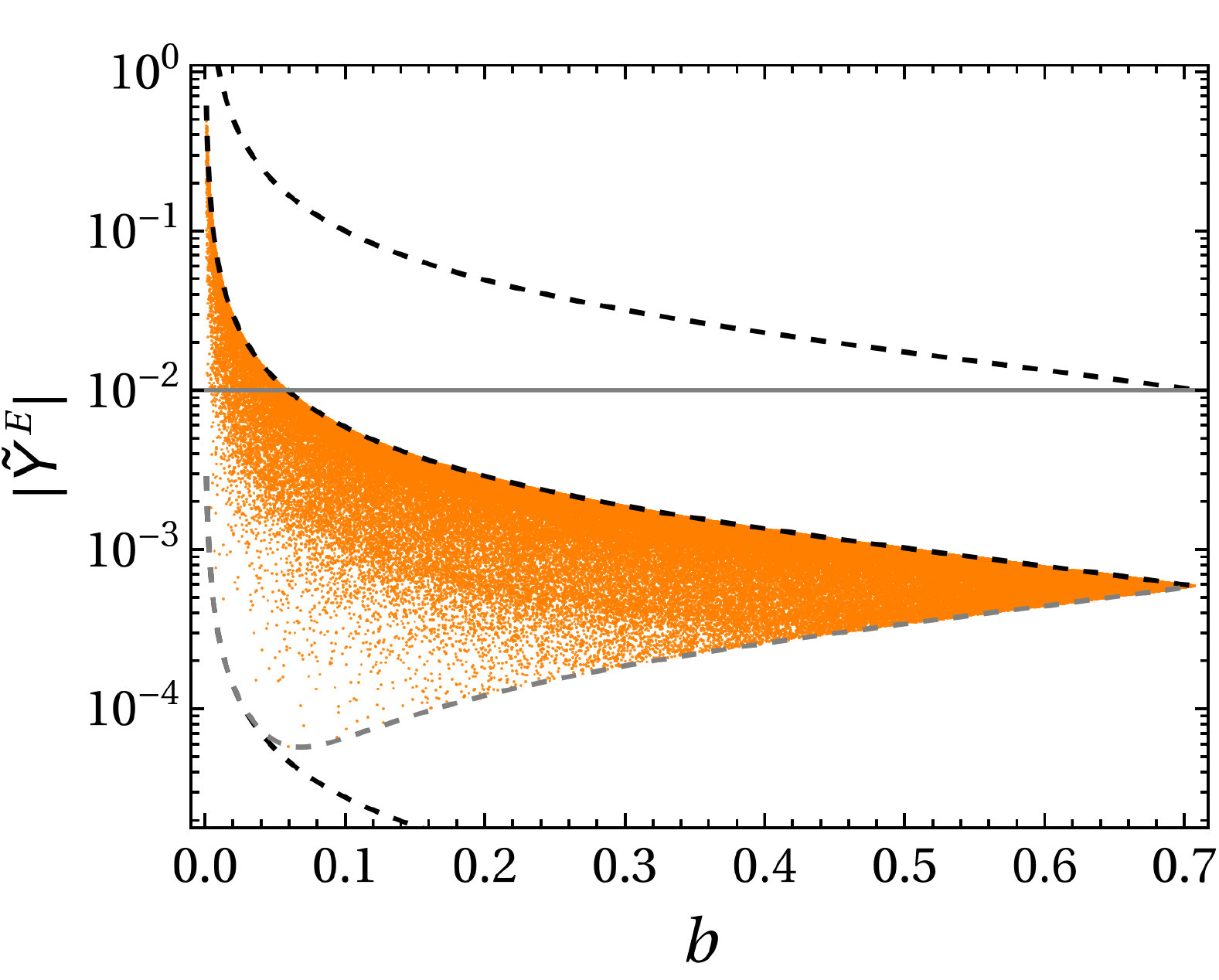}
\caption{\label{fig:333}
Special case where third row of the PMNS is real (by choosing $\beta_1,\beta_2$).
\textbf{Left}: the Yukawa couplings $|\tY^E_1|,|\tY^E_2|$ as a function of $b$ ($\tY^E_3=0$).
Other settings are the same as Fig.\,\ref{fig:Y-tilde}.
\textbf{Right}: the norm $|\tY^E|$ as a function of $b$.
The continuous gray line is $y_\tau$. The dashed black lines are $y_i/b$, $i=e,\mu,\tau$ while the dashed gray line is the lower bound function in \eqref{YEmax:norm} with $y_{\rm max}=y_\mu$ and $y_{\rm min}=y_e$.
Other settings are the same as Fig.\,\ref{fig:Y-norm}.
}
\end{figure}

Although all the patterns in \eqref{tYE:special} have one vanishing component, the third pattern with $\tY^E_3=0$ has suppressed $|\tY^E|$ compared to the other cases.
For the latter, and generically, the norm $|\tY^E|$ is approximately bounded from above by $y_\tau/b$; see Fig.\,\ref{fig:Y-norm}. In contrast, for the special case where $\tY^E_3=0$, the upper bound is decreased to  $y_\mu/b$, i.e., 17 times smaller.
For illustration, we show in Fig.\,\ref{fig:333} the components $|\tY^E_1|,|\tY^E_2|$ (left) and the norm $|\tY^E|$ (right) as a function of $b$ for the case where the third row of the PMNS ($u_3$) is real;
$|\tY^E_3|=0$ is not shown.
Comparing to Figs.\,\ref{fig:Y-tilde} and \ref{fig:Y-norm}, we can see that the Yukawa couplings are very suppressed compared to the typical value $y_\tau/\sqrt{3}$ in the gray continuous line.
See appendix \ref{ap:seesaw} for details.

For the rest of this subsection, we prove the converse: if $\tY^E_i=0$ independently of $b$, then a eigenvector $u_i$ is real up to rephasing.

To prove that, let us name the columns of the real orthogonal matrix $\cO$ in \eqref{def:cO}:
\eq{
\label{cO:ei}
\cO= \left(
    \begin{array}{c|c|c}
    e_1 & e_2 & e_3
    \end{array}
    \right)\,.
}
We also denote the first vector as $e_0=e_1$ because it is the eigenvector of zero eigenvalue of $A_1^{-1/2}A_2A_1^{-1/2}$. The vectors $e_2,e_3$ span the orthogonal space.

Now, using Eqs.\,\eqref{formula:cal-Yd} and \eqref{w}, we can write $Y^E$ in \eqref{YE:NB} as
\eq{
Y^E=x_aA_1^{1/2}e_3+ix_bA_1^{1/2}e_2\,,
}
where $x_a=a/\sqrt{1-a^2}$ and similarly for $x_b$. Note that they are not independent as $x_ax_b=\mu$ from \eqref{def:mu}.

To calculate the $j$-th component of $\tY^E$, we can project on the $j$-th vector of \eqref{VeL:ui}:
\eq{
\tY^E_j=(V_{e_L}^\dag Y^E)_j=x_au_j^\dag A_1^{1/2}e_3+ix_b u_j^\dag A_1^{1/2}e_2\,.
}
Vanishing of this component independently of $b$ requires
\eq{
u_j^\dag A_1^{1/2}e_3=u_j^\dag A_1^{1/2}e_2=0\,.
}
This implies that $A_1^{1/2}u_j$ is orthogonal to both $e_2$ and $e_3$, hence it should lie in the direction of $e_1=e_0$.
Because $A_1^{1/2}$ is real and nonsingular, both the real and imaginary part of $u_j$ should lie in the direction of $A_1^{-1/2}e_0$. Thus $u_j$ is real up to rephasing. But then, from the discussion of the first part, $u_j$ is also an eigenvector of $A_1$ similarly to \eqref{real.u3} and $u_j=e_0$ up to rephasing.

\subsection{NB-VLL decoupling of two flavors}

We can now ask if instead of one vanishing component for $\tY^E_i$, two vanishing components are possible.
We will show that the following patterns are possible:
\eq{
\label{tYE:special:2}
(\tY^E_e,\tY^E_\mu,\tY^E_\tau)=(\times,0,0)
\text{~~or~~}
(0,\times,0)\,.
}
The case $\tY^E_e=\tY^E_\mu=0$ is not possible.
For these patterns, \emph{there is no} flavor changing neutral current coupling to $Z$ at tree-level.
We will also conclude that these patterns only arise as subcases of the patterns discussed in Sec.\,\ref{sec:1flavor=0} and occur only for special values of $b$ and $\gamma$, or equivalently the direction $e_2$ in \eqref{cO:ei}.
For the best-fit values of mixing angle and $\delta$ we show the possible cases in table \ref{tab:special}.
The convention for the betas, each in the range $(-\pi/2,\pi/2)$, follow \eqref{pmns} and note the periodicity in $\beta_i\to \beta_i+\pi$.
\begin{table}[h]
\eq{\nonumber
\begin{array}{|c|c|c|c|c|c|}
\hline
\text{Pattern $\tY^E_i$} & (\beta_1,\beta_2) & b & e_2 & |\tY^E| & e_0 \text{ direction}
\\
\hline 
(\times,0,0) & (0.21270, -0.74054) & 0.14397 & 
    \left(
    \begin{array}{c}
    0.30196 \\
    -0.60405 \\
    0.73753 \\
    \end{array}
    \right) 
    & y_\mu|w|\sim 5.9\times 10^{-4} & u_3~\text{real}
\\
\hline
(0,\times,0) & (0,0) & 0.42283 & 
    \left(
    \begin{array}{c}
    0.19314 \\
    -0.02394 \\
    -0.98088 \\
    \end{array}
    \right)
    & y_\tau|w|\sim 10^{-2} & u_1~\text{real}
\\
\hline
(\times,0,0) & (-0.29900, -1.1009) & 0.18484 & 
    \left(
    \begin{array}{c}
    0.32304 \\
    0.79472 \\
    0.51387 \\
    \end{array}
    \right)
    & y_\tau|w|\sim 10^{-2} & u_2~\text{real}
\\
\hline
\end{array}
}
\caption{\label{tab:special}
Special points with two vanishing $\tY^E_i$. PMNS mixing angles and $\delta$ are fixed at best-fit\,\cite{capozzi:global}. $u_j$ real means that the $j$-th row of PMNS is real.
}
\end{table}

For the proof, taking $\tY^E$ with only one non-vanishing component means that $Y^E$ is in the direction of some eigenvector $u_i$ of $Y^e{Y^e}^\dag$; see \eqref{YYe:ui}.
This means that
\eq{
Y^e{Y^e}^\dag Y^E=y_i^2Y^E\,.
}
Substituting Eqs.\,\eqref{YE:NB} and \eqref{Ye:NB:S}, we obtain
\eq{
\label{calY.w}
{\cY^e}^\tp \cY^e w=y_i^2(\id_3-ww^\dag)^{-1}w=\frac{y_i^2}{1-|w|^2}w\,.
}
The last equality follows because $w$ is an eigenvector of $(\id_3-ww^\dag)$.
Using \eqref{w}, we conclude that ${\cY^e}^\tp \cY^e$ needs to be diagonal with two degenerate eigenvalues:
\eq{
\label{calY:ortho}
{\cY^e}^\tp \cY^e=\diag\left(\times,\frac{y_i^2}{1-|w|^2},\frac{y_i^2}{1-|w|^2}\right)\,.
}
This also implies that the columns of $\cY^e$ are orthogonal vectors.

Next, we use the explicit parametrization \eqref{formula:cal-Yd} in \eqref{calY:ortho}. We obtain that 
\eq{
\label{A1.diag}
\cO^\tp A_1\cO=\diag\left(\times,\frac{y_i^2}{1-|w|^2}(1-b^2),\frac{y_i^2}{1-|w|^2}(1-a^2)\right)\,,
}
is also diagonal. The unknown entry is the same as in \eqref{calY:ortho}.
Because of \eqref{A1.diag}, and from the definition in \eqref{def:cO}, the matrix $\cO$ transforms $A_2$ into canonical form as well:
\eq{
\cO^\tp A_2\cO=\mtrx{0 &&\cr &0&-\mu'\cr &\mu'&0}\,.
}
The overall parameter is $\mu'=\mu\frac{y_i^2}{1-|w|^2}\sqrt{1-b^2}\sqrt{1-a^2}$.

Since $A_1+iA_2$ is $Y^e{Y^e}^\dag$, we obtain
\eq{
\label{YYe.block}
\cO^\tp Y^e{Y^e}^\dag\cO=
\diag(1,\sqrt{1-b^2},\sqrt{1-a^2})
\mtrx{\times &\cr & \frac{y_i^2}{1-|w|^2}\mtrx{1&-i\mu\cr i\mu&1}}
\diag(1,\sqrt{1-b^2},\sqrt{1-a^2})\,.
}
Since this matrix is already block diagonal, the unknown entry should be an eigenvalue $\times=y_j^2$, $j\neq i$.
This is the same entry in \eqref{calY:ortho} and \eqref{A1.diag}.
Moreover, we identify the first column of $\cO$ as a common eigenvector of $Y^e{Y^e}^\dag$: $e_0=u_j$.
Thus $u_j$ is real up to rephasing.
Explicit diagonalization of the $2\times 2$ subblock in \eqref{YYe.block} leads to the following spectrum for 
\eq{
Y^e{Y^e}^\dag:\quad \left(y_j^2,\,y_i^2,\,\frac{y_i^2}{1-|w|^2}\right)\,.
}
Since $|w|<1$, correspondence to $(y^2_e,y^2_\mu,y^2_\tau)$ is only possible for $(j,i)=(3,1), (2,1), (1,2)$.
These possibilities cover all the cases in table \ref{tab:special}.
We also see that the spectrum of ${\cY^e}^\tp \cY^e$ in \eqref{calY:ortho} contains only two of the eigenvalues of $Y^e{Y^e}^\dag$.

As a bonus, from the relation \eqref{calY.w}, we obtain the norm of $Y^E$ in \eqref{YE:NB} as
\eq{
|Y^E|^2=\frac{y_i^2|w|^2}{1-|w|^2}=y_k^2|w|^2\,,
}
where $k\neq i,j$.

%%%%%%%%%%%%%%%%%%%%%%%%%%%%%%%%%%%%%%%%%%%
\section{Parameter space}
\label{sec:params}

Considering the LEP bound\,\cite{LEP} for the pair production of VLLs, we take conservatively
\eq{
\label{LEP:bound}
M_E\ge 100\,\unit{GeV}\,.
}
Although fairly model independent, the pair production cross section of charged singlet VLLs is not particularly large and direct searches at LHC have only put mild constraints\,\cite{vll.lhc}.\,%
\footnote{A recent reanalysis can be seen in Ref.\,\cite{vll.lhc:recast}.}

The flavored Yukawa couplings $Y^E_i$ of the VLL to the SM induce deviations on the charged current to $W$ and the neutral current to $Z$. The flavor conserving part of the latter are strongly constrained by EW observables. 
A global fit including these observables and lepton flavor universality tests roughly constrains\,\cite{crivellin:global}\,\footnote{%
We roughly projected on the 95\% CL region with 2 dgf. We took the more conservative case that includes the ``new nuclear corrections'' that modify the prediction for $|V_{ud}|$.}
\eq{
\label{tYE:bound:0}
\frac{v}{M_E}|\tY^E_1|<0.04\,,\quad
\frac{v}{M_E}|\tY^E_2|<0.035\,,\quad
\frac{v}{M_E}|\tY^E_3|<0.05\,.
}
The last value translates into $|\Theta_{3}|<0.035$ which would be compatible with $|V_{E2}|<0.034$ found in Ref.\,\cite{delAguila} (the other components are smaller); see also Ref.\,\cite{strumia}.
For simplification, we will consider the last value in \eqref{tYE:bound:0} as a bound on all components.
Then, we can translate the bounds \eqref{tYE:bound:0} into a bound on each Yukawa coupling:
\eq{
\label{tYE:bound}
|\tY^E_i|< 0.02\times \frac{M_E}{100\,\unit{GeV}}\,.
}
This bound tell us that we need to worry about non-perturbative values for $|Y^E_i|$ only for $M_E\gtrsim 50\,\unit{TeV}$.
We show the bound \eqref{tYE:bound} for two masses as dashed lines in Fig.\,\ref{fig:Y-tilde}.
The lines for other masses can be read by linear scaling.

The bounds \eqref{tYE:bound:0} also allows us to quantify the maximum deviation $\delta M_E$ in the VLL mass coming from the full diagonalization of \eqref{mass.matrix} (right) compared to the input mass $M_E$:
\eq{
\label{seesaw.dev}
\frac{\delta M_E}{M_E}\lesssim \frac{v^2}{4M_E^2}\tY^E{}^{\dag} \tY^E\approx 0.2\%\,.
}
This value also approximately quantifies the quality of the leading seesaw approximation \eqref{leading.ss}.
See appendix \ref{ap:seesaw} for more details.

We can now discuss the bounds coming from the CLFV processes.
We have discussed in Sec.\,\ref{sec:current.lim} how CLFV processes constrain typical values of the Yukawa couplings $\tY^E_i$ of the NB-VLL with the SM and have seen that the no observation of $\mu\to e$ flavor change leads to particularly severe constraints.
However, as we can be seen in Fig.\,\ref{fig:Y-tilde}, these Yukawa couplings can vary over a wide range.
Moreover, the possibility of vanishing Yukawa couplings in special points of parameter space, cf.\,Sec.\,\ref{sec:special}, does not allow us to put strict constraints for specific flavor changing channels.
Nevertheless, we will show in the following that except for points near these special points, the constraint from the $\mu\to e$ conversion in nuclei pushes the VLL mass to at least hundreds of GeV.

\begin{figure}[h]
\includegraphics[scale=0.5]{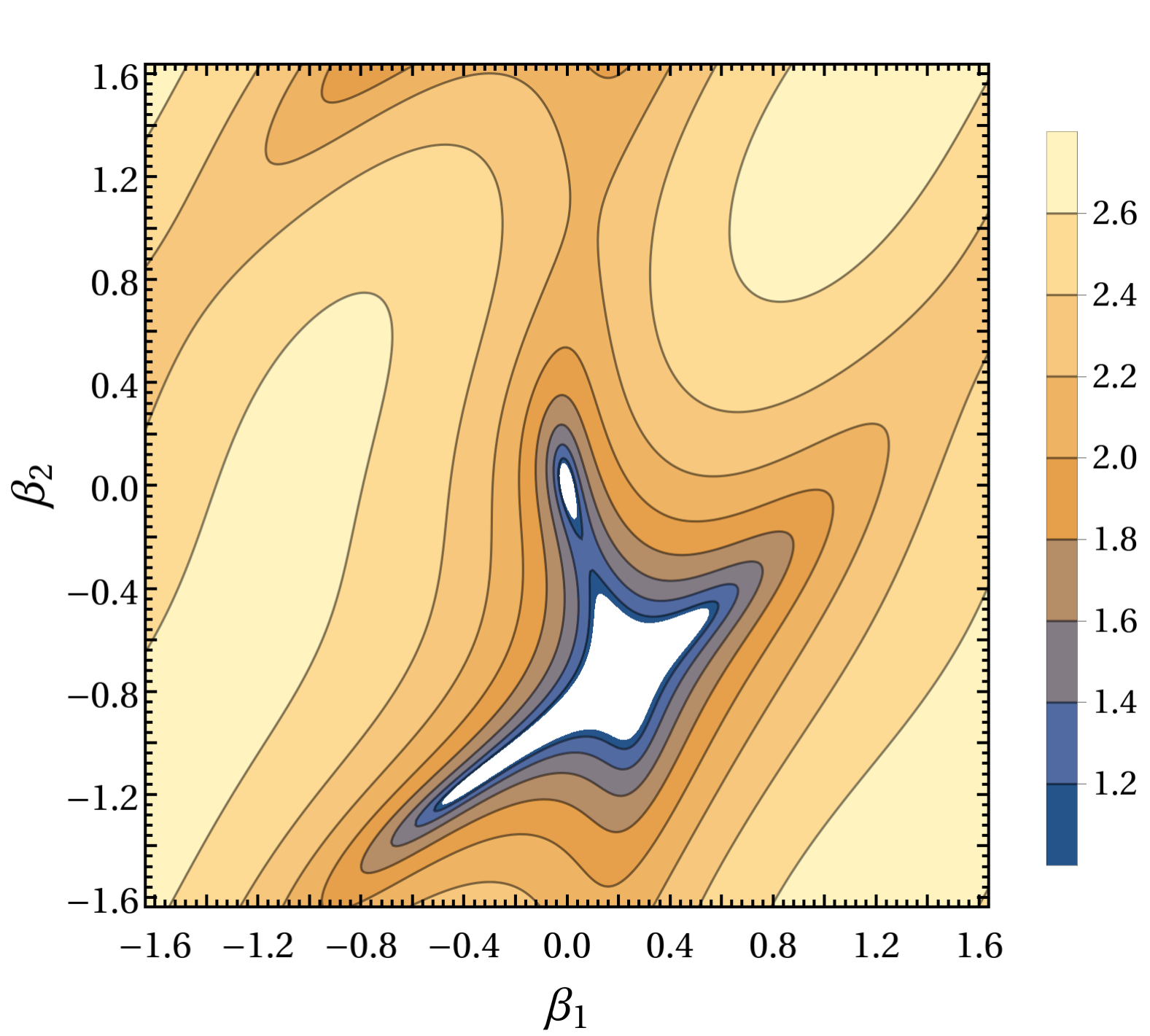}
\includegraphics[scale=0.5]{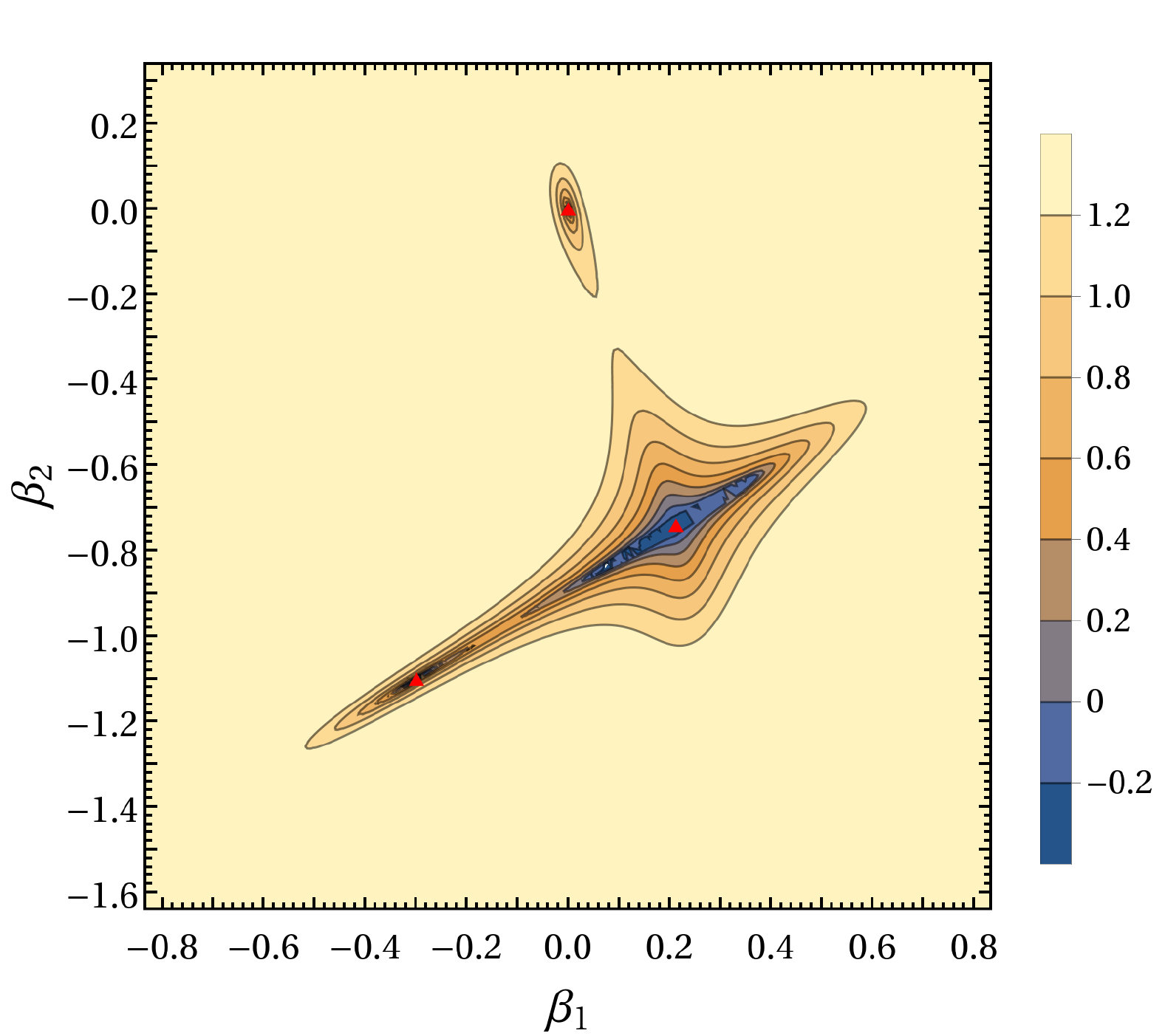}
\caption{\label{fig:beta12}%
Contour plot of $\log_{10}(10^{+7}|\tY^E_e\tY^E_\mu|)$ in the plane of Majorana phases $(\beta_1,\beta_2)$ in the whole range $[-\pi/2,\pi/2]$ (\textbf{left}) and around the three minima (\textbf{right}); cf.\,table\,\ref{tab:special}.
The PMNS matrix parametrized in the convention \eqref{pmns} is fixed at the best-fit values for $\theta_{ij}$ and $\delta=1.28\pi$ of Ref.\,\cite{capozzi:global}.
}
\end{figure}

To organize the extraction of constraints, let us define from \eqref{constraint:Yemu} the exponent
\eq{
\label{def:log}
10^{k}\equiv 10^{+7}|\tY^E_\mu\tY^{E*}_e|< \left(\frac{M_E}{142\,\unit{GeV}}\right)^2\,.
}
We then show in Fig.\,\ref{fig:beta12} the exponent $k$ of the lefthand side in the plane of Majorana phases $(\beta_1,\beta_2)$ minimized with respect to the rest of parameters ($b,\gamma$) keeping angles and the CP phase of the PMNS in their best-fit values of Ref.\,\cite{capozzi:global}.
We can see that every point except the white region ($k<1$) requires that
\eq{
\label{most.ME.limit:current}
M_E>\sqrt{10}\times 142\,\unit{GeV}=449\,\unit{GeV}\,.
}
The region inside the highest contour of value $k=2.6$ requires $M_E>2.8\,\unit{TeV}$.
For comparison, if we use the reference value \eqref{tYE.ref}, we obtain $k=2.52$.
In the right panel, which focuses on the white region, we clearly see the special points of table~\ref{tab:special} which are marked by red triangles.

\begin{figure}[h]
\includegraphics[scale=0.5]{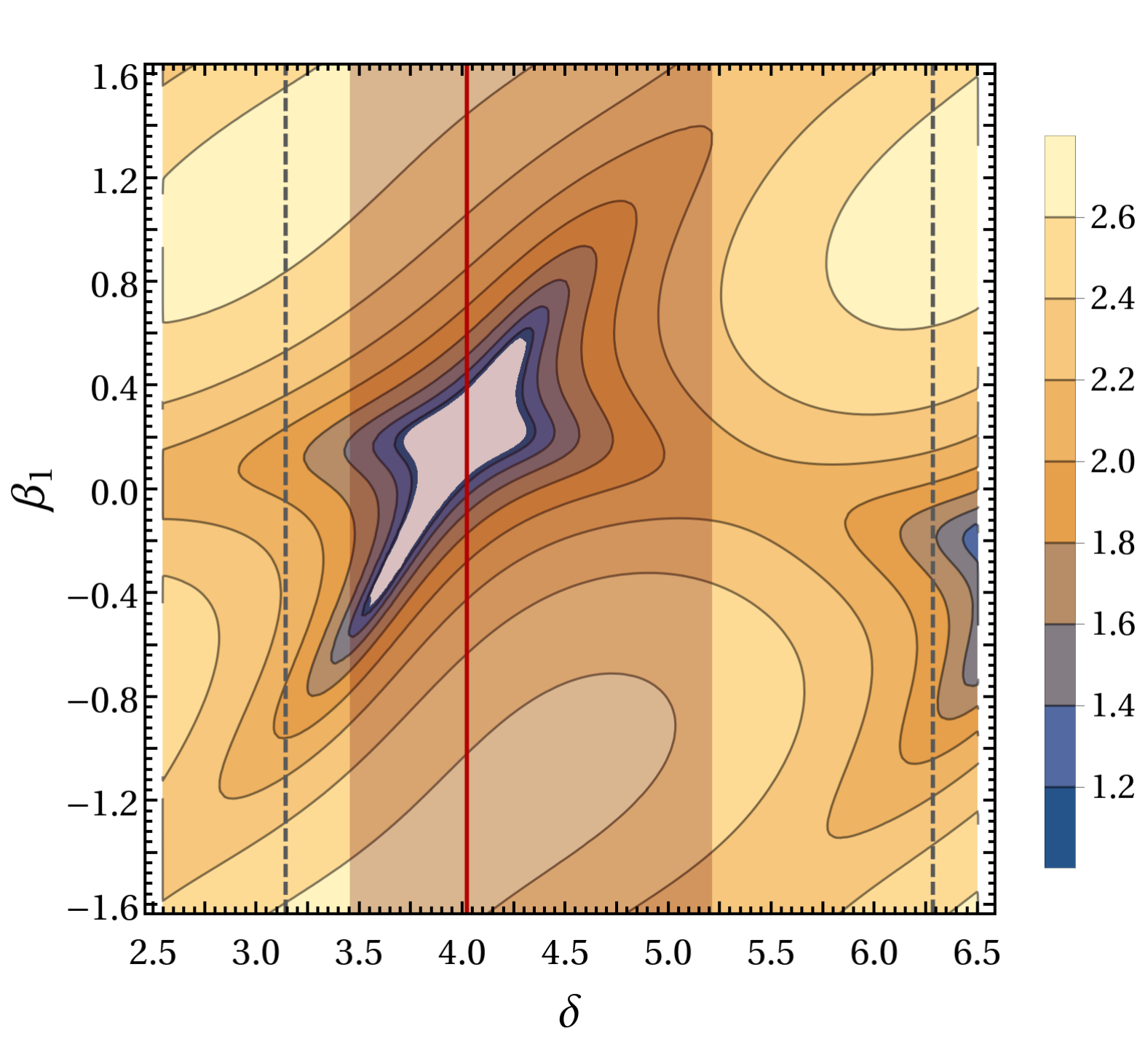}
\includegraphics[scale=0.5]{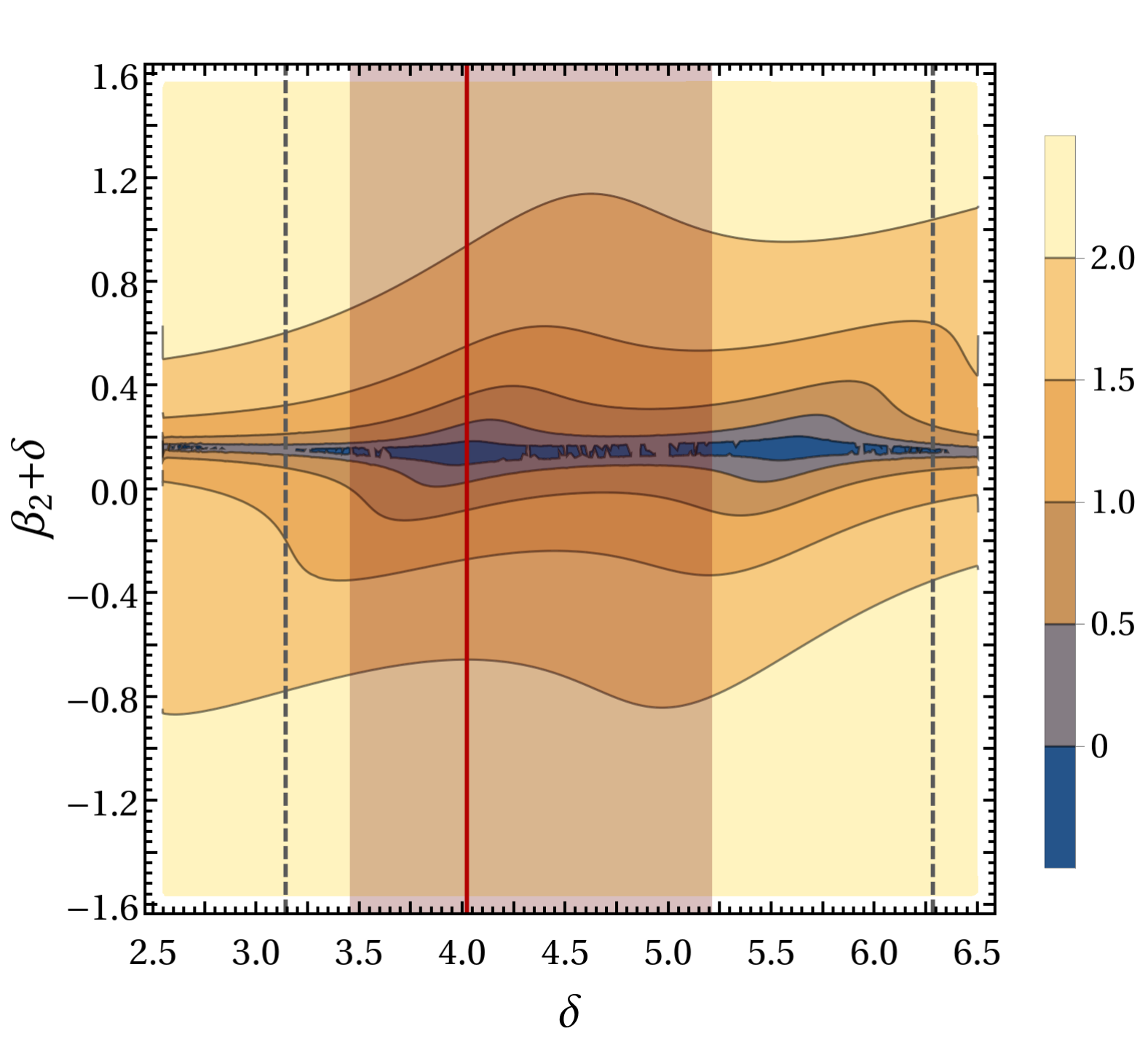}
\caption{
\label{fig:beta-delta}
Contour plot of $\log_{10}(10^{+7}|\tY^E_e\tY^E_\mu|)$ in the plane $(\delta,\beta_1)$ [\textbf{left}] and $(\delta,\beta_2+\delta)$ [\textbf{right}].
The Majorana phase $\beta_1$ or the combination $\beta_2+\delta$ are varied in the whole range $[-\pi/2,\pi/2]$ while the Dirac CP phase $\delta$ is varied within the 3$\sigma$ range of Ref.\,\cite{capozzi:global} for NO: $[0.81\pi,2.07\pi]$. Usual periodicity is assumed. The red line marks the best-fit value of $\delta=1.28\pi$ and the dashed gray lines mark the CP conserving values $\delta=\pi,2\pi$.
}
\end{figure}

The contours in Fig.\,\ref{fig:beta12} do not change much if PMNS mixing angles are varied within their $3\sigma$ values. But expectedly they change substantially when the CP phase $\delta$ is varied.
To illustrate this change, we show in Fig.\,\ref{fig:beta-delta} the same quantity $k$ in the planes $(\delta,\beta_1)$ [left] and $(\delta,\delta+\beta_2)$ [right] keeping the PMNS mixing angles the same. 
The Majorana phase $\beta_1$ and the combination $\delta+\beta_2$ are varied in the interval $[-\pi/2,\pi/2]$ while $\delta$ is varied in the $3\sigma$ range $[0.81\pi,2.07\pi]$ of Ref.\,\cite{capozzi:global}, assuming the usual periodicity.
In the left panel, $\beta_2=-0.74$ is fixed to the value of the first special point of table \ref{tab:special} where the third row of the PMNS is real.
In the right panel, $\beta_1=0.21$ is equally fixed to the value of the same special point.
The red lines mark the best-fit value for $\delta$ while the shaded bands mark the $1\sigma$ interval.
The CP conserving values $\delta=\pi,2\pi$ are also shown in dashed lines to aid the visualization.

All plots show that the region around the special points have suppressed $|\tY^E_e\tY^E_\mu|$.
In the plane $(\beta_1,\beta_2)$ and $(\delta,\beta_1)$ these regions are localized.
In the plane $(\delta,\delta+\beta_2)$ the region of suppressed $|\tY^E_e\tY^E_\mu|$ is not localized but extends over a band of fixed value for $\delta+\beta_2\approx 0.14$ corresponding to the value where the third row of the PMNS is approximately real up to rephasing. 
We note that Fig.\,\ref{fig:beta-delta} assumes specific values for the fixed betas and different contours are obtained for different values.
Nevertheless, the regions of suppressed $|\tY^E_e\tY^E_\mu|$ are always confined to small regions and the larger values are roughly the same.
We also emphasize that most of these minimal values corresponding to the contours are usually much smaller than the typical values for the charged NB-VLL. For typical values, a large part of the parameter space is excluded from $\mu\to e$ conversion in nuclei.

In the future, the Mu2e experiment\,\cite{mu2e} is expected to dramatically improve the limit to the one in \eqref{mu2e.limit}.
This limit can be written in terms of $k$ as
\eq{
10^{k} < \left(\frac{M_E}{870\,\unit{GeV}}\right)^2\,.
}
So the limit \eqref{most.ME.limit:current} will be improved to
\eq{
\label{most.ME.limit:future}
M_E>\sqrt{10}\times 870\,\unit{GeV}=2.75\,\unit{TeV}\,,
}
for all points in Fig.\,\ref{fig:beta12} excluding the white region.

%%%%%%%%%%%%%%%%%%%%%%%%%%%%%%%%%%%%%%%%%%
\section{Conclusions}
\label{sec:conclu}

We have studied the flavor constraints for a model where all the leptonic CP violation in the SM were originated from the soft CP violation ---quantified by a single CP phase--- in the mixing term of one singlet charged vector-like lepton with the SM charged leptons.
Then, if neutrinos are Majorana, the three CP violating phases residing in the PMNS mixing should be induced by this single phase.
Since the mixing term is the single source of CP violation, the vector-like lepton cannot decouple from the SM if leptonic CP violation is confirmed to exist in nature.
This kind of setting may arise from models implementing the Nelson-Barr scheme to solve the strong CP problem where vector-like quarks transmit the CP violation that leads to the CKM phase without inducing a $\bar{\theta}$ parameter at tree level.
So we have dubbed these vector-like leptons as NB-VLLs.

In contrast to the quark case, the Yukawa couplings of the vector-like lepton with the SM are not hierarchical due to the large mixing angles of the leptonic sector 
and significant deviation from MFV is allowed.
Therefore, typical values of these Yukawa couplings are severely constrained by charged lepton flavor violating processes, specially the $\mu\to e$ conversion in nuclei, pushing the mass scale of the NB-VLL to the TeV range.

These conclusions are, however, not applicable to the entirety of the parameter space.
We have identified special points where the NB-VLL can decouple from one or even two lepton flavors and all these cases were mapped; see table \ref{tab:special}.
In the case of decoupling of two flavors, many flavor violating observables vanish and the induced constraints are obviously avoided.
These special points, however, are highly sensitive to the Majorana phases of the PMNS matrix and correspond to cases where the \emph{rows} of the latter are real.
Therefore, these points are highly fine-tuned and thus highly unlikely to occur unless they can be protected by symmetry.

By continuity, the small regions close to these special points have suppressed Yukawa couplings to the NB-VLL, specially if the latter is decoupled from the $\tau$ flavor.
These regions can be seen in Figs.\,\ref{fig:beta12} and \ref{fig:beta-delta}.
Excluding these small regions, the mass of the NB-VLL is guaranteed to be above 450\,GeV, cf.\,\eqref{most.ME.limit:current}. Future experiments on $\mu\to e$ conversion in nuclei will improve these limit to 2.7\,TeV, cf.\,\eqref{most.ME.limit:future}.
Therefore, the combination of direct detection of VLLs in colliders and low energy flavor changing experiments can in principle rule out a single charged NB-VLL for certain parts of the parameter space.
More precise predictions can be made once we measure the Dirac CP phase of the leptonic mixing.

In summary, the possibility of generating leptonic CP violation from the mixing of vector-like leptons is a very interesting possibility that can be tested in the near future and may be linked to more fundamental questions connected to the origin of CP violation in nature.

%%%%%%%%%%%%%%%%%%%%%%%%%%%
\acknowledgments

C.C.N.\ acknowledges partial support by Brazilian Fapesp, grant 2014/19164-6, and
CNPq, grant 304262/2019-6. 
G.D.C.\ acknowledges financial support by the Coordenação de Aperfeiçoamento de Pessoal de Nível Superior - Brasil (CAPES) - Finance Code 001.

%%%%%%%%%%%%%%%%%%%%%%%%%%%%%%%%%%%%%%%%%%%
\appendix
%%%%%%%%%%%%%%%%%%%%%%%%%%%%%%%%%%%%%%%%%%%
\section{Seesaw deviation and approximate bounds}
\label{ap:seesaw}

We can quantify the deviation from the leading seesaw approximation by using the next order correction to the VLL mass.
For one singlet VLL, the deviation is
\eq{
\label{deltaME}
\frac{\delta M_E}{M_E}\approx \frac{\delta M_E^2}{2M_E^2}=\ums{2}|\theta_L|^2
=\frac{v^2}{4M_E^2}|\tY^E|^2
\,,
}
where $\theta_L\sim 3\times 1$ is given in \eqref{thetaL} and the factor half is due to the square root.
Note that $|\tY^E|=|Y^E|$.
Therefore the bound \eqref{tYE:bound} translates to \eqref{seesaw.dev}.
This quantification is valid for a generic charged singlet VLL including the case of a NB-VLL.

Specifically for the singlet NB-VLL case, our parametrization relations \eqref{formula:cal-Yd} together with \eqref{YE:NB} allows to write
\eqali{
\label{YEnorm}
|\tY^E|^2&={Y^E}^\dag Y^E = w^\dag \cO^\tp \re(Y^e{Y^e}^\dag)\cO w\,,
\cr
&= \re(\cO^\tp Y^e{Y^e}^\dag\cO)_{22}x_b^2 + \re(\cO^\tp Y^e{Y^e}^\dag\cO)_{33}x_a^2\,,
}
where $x_b\equiv b/\sqrt{1-b^2}$ and analogously for $x_a$. The latter are not independent and are related by $x_a=\mu/x_b$ with $\mu$ typically unity.\footnote{For example, $1-\mu$ varies in the range $2\times 10^{-7}\text{ -- }10^{-2}$ for best-fit mixing angles and $\delta$.}
Considering $b<a$ and assuming generic $\cO$, we obtain the simple bound for the norm as a function of $b$,
\eq{
\label{YEmax:norm}
|\tY^E|\lesssim y_\tau\frac{\mu}{x_b}\approx 10^{-2}\times\frac{\mu}{x_b}\,,
}
where $y_\tau$ is the $\tau$ Yukawa.
This relation can be used to set a lower value for $b$ to guarantee a certain maximal deviation \eqref{deltaME} for the seesaw approximation. Similar relations are valid for vector-like quarks of NB type \cite{nb-vlq,nb-vlq:fit}.

Similarly, we can find upper bound functions for the components $Y^E_i$ from
\eq{
Y^E=\left[\re\bigg(Y^e{Y^e}^\dag\bigg)\right]^{1/2}\cO \mtrx{0\cr ix_b\cr \dst\frac{\mu}{x_b}}\,.
}
But differently from the quark case, the matrix $Y^e{Y^e}^\dag$ is not hierarchical.
Therefore the components may take similar values.
Approximating $\re(Y^eY^e{}^\dag)$ by $Y^eY^e{}^\dag$ and taking the norm of each row because of $\cO$, we get
\eqali{
\label{YEimax}
|Y^E_i|&\lesssim \frac{\mu}{x_b}\times 10^{-3}\times (4.62, 5.81, 6.68)\,,
}
using best-fit values for the mixing angles and $\delta$.
The expression \eqref{YEimax} gives a good estimate of the maximal values taken by $|Y^E_i|$ as a function of $b$ and is very similar for NB-VLQs\,\cite{nb-vlq:fit}.
Note that the upper limits in \eqref{YEimax} are close to $y_\tau \mu/(\sqrt{3}x_b)$ and they respect \eqref{YEmax:norm}. 
These expressions are approximately valid for $|\tY^E_i|$ but simple relations are harder to find. Moreover, as seen in Sec.\,\ref{}, individual $|\tY^E_i|$ may vanish for special regions of the parameter space.
Then it is clear that lower bounds are inexistent for individual $|\tY^E_i|$.

In contrast, a lower bound can be found for the norm \eqref{YEnorm}. Let us recall that the real orthogonal matrix $\cO$ is defined by \eqref{def:cO}. In this definition, the first column of $\cO$ is given by an eigenvector $e_0$ with zero eigenvalue while the rest of the columns span the orthogonal space.
Let us denote by $P_\perp$ the projector onto this orthogonal space.
Then the expression in \eqref{YEnorm} is bounded by
\eqali{
\label{YEnorm:bounds}
|Y^E|^2_{\rm max}&=y_{\rm min}^2x_b^2+y_{\rm max}^2x_a^2\,,\cr
|Y^E|^2_{\rm min}&=y_{\rm max}^2x_b^2+y_{\rm min}^2x_a^2\,,\cr
}
where $y^2_{\rm max},y^2_{\rm min}$ are given respectively by the largest and smallest eigenvalue of $P_\perp \re\bigg(Y^e{Y^e}^\dag\bigg)P_\perp$, excluding the zero eigenvalue.
These expressions follow from the extremization of $x^\tp Bx$ in two dimensions with $x$ restricted to the circle and $B$ positive definite.
The accuracy of these bounds may be checked numerically.

We can now specialize to the situations described in Sec.\,\ref{sec:1flavor=0} where one of the components $\tY^E_j=0$ and the corresponding $j$-th row of the PMNS matrix is real.
This is equivalent to $Y^E\sim u_j=e_0$, where these vectors are defined in \eqref{VeL:ui} and \eqref{cO:ei}.
From the property \eqref{real.u3} and subsequent discussion, the matrix $\cO^\tp Y^e{Y^e}^\dag\cO$ is block-diagonal: 
\eq{
\label{YYe.cO}
\cO^\tp Y^e{Y^e}^\dag\cO=\mtrx{y^2_j&&\cr &\times&\times\cr &\times&\times}\,.
}
The diagonal elements of the lower $2\times 2$ subblock enters the expression in \eqref{YEnorm}.
Because the eigenvalue $y^2_j$ is singled out in \eqref{YYe.cO}, the lower subblock only depends on the other eigenvalues $y^2_i,y^2_k$, $i\neq j\neq k$, and these correspond to the values $y^2_{\rm min},y^2_{\rm max}$ in \eqref{YEnorm:bounds} if we do not take the real part in \eqref{YEnorm}.
With the real part, for a complex $Y^e{Y^e}^\dag$, the pair $(y^2_{\rm min},y^2_{\rm max})$ will correspond to the eigenvalues of the subblock which will necessarily lie inside the interval $(y^2_i,y^2_k)$.
We can think that the eigenvalues move with increasing complex part.
For a hierarchical spectrum, we have checked that the largest eigenvalue moves very little compared to the smallest eigenvalue and we can approximate $y^2_{\rm max}\approx \max(y^2_i,y^2_k)$.

The case where $y^2_j=y^2_3$ is then special because $y_{\rm max}\approx y_2=y_\mu$ and
\eq{
|Y^E|\lesssim y_\mu\frac{\mu}{x_b}\approx 5.9\times 10^{-4}\frac{\mu}{x_b}\,.
}
For the best-fit values of angles and $\delta$ of PMNS, $y_{\rm min}=8.5\times 10^{-5}\gg y_e$ and $|Y^E|_{\rm min}=3.1\times 10^{-4}$.
If we vary the mixing angles and $\delta$, we can adopt $y_{\rm min}=y_e$; see Fig.\,\ref{fig:333}.
In this case the minimal value is $|Y^E|_{\rm min}=\sqrt{2y_ey_\mu}=5.7\times 10^{-5}$ for $x_b^2=y_e/y_\mu$.

%%%%%%%%%%%%%%%%%%%%%%%%%%%%%%%%%%%%%%%%%%%
\section{Radiative decay formula}
\label{ap:formulas}

We present here the full contribution to the dipole operator \eqref{dipole} coming from the presence of one charged singlet VLL, subtracted from the SM contribution.
\eqali{
\label{cR:loop}
c_R^{\ell_f\ell_i}&=\frac{e}{16\pi^2}\sqrt{2}G_F m_{\ell_i}\left\{
X_{Ef}^*X_{Ei}\left[
\tF_V(x_E)+y_E\Big(\tF_S(y_E)+F_S(y_E)\Big)
\right]
\right.
\cr
&\quad 
+\big[(X_{fi}-\delta_{fi})(1-4s_w^2)-X^*_{Ef}X_{Ei}\big]\tF_V(0)
\cr
&\quad +
\left.
(X_{fi}-\delta_{if})\big[2\tf_V(0)-2s_w^2F_V(0)\big]
\right\}\,.
}
We have neglected the SM lepton masses in the loop.
The expression is still valid for large mixings and can be simplified in the seesaw approximation using 
$X_{fi}-\delta_{fi}\approx -X_{Ef}^*X_{Ei}$.
Within this approximation, the last two lines result in
\eq{
\label{rad:cte}
X^*_{Ef}X_{Ei}\big[-2\tf_V(0)+2s_w^2F_V(0)
-2(1-2s_w^2)\tF_V(0)\big]
=X^*_{Ef}X_{Ei}\left(\frac{1}{6}-\frac{2}{3}s^2_w\right)\,,
}
with the numerical factor being $\frac{1}{6}-\frac{2}{3}s^2_w=0.0125$.
The loop functions were listed in Sec.\,\ref{sec:muegamma} and their numerical values at zero are
$\tf_V(0)=-\frac{5}{12}, F_V(0)=-1, \tF_V(0)=\frac{1}{3}$.

The coefficient above leads to the expression \eqref{Br:mu.e.gamma} for the branching of $\mu\to e\gamma$ which differs from Ref.\,\cite{crivellin:global} where the contribution above depending on $F_S(y_E)$ and $F_V(0)$ are missing; see Sec.\,\ref{sec:muegamma}.
These contributions involve righthanded vertices and the latter involves a diagram with ordinary leptons and $Z$ in the loop.
Without this term, the factor inside the parenthesis in \eqref{rad:cte} becomes $\frac{1}{6}+\frac{4}{3}s^2_w= 0.475$, which is 38 times larger.

In the seesaw approximation, the factor between curly brackets in \eqref{cR:loop}, with $X^*_{Ef}X_{Ei}$ factored, is a monotonically decreasing function of $M_E$, decreasing from 0.11 to $-0.064$ when $M_E$ increases from $100\,\unit{GeV}$ to 1\,TeV. At much larger masses, it asymptotically approaches $-0.071$.
The zero of the function occurs at $M_E\approx 239\,\unit{GeV}$ and it is beyond this mass that the contribution to $a_\mu$ becomes positive.

%%%%%%%%%%%%%%%%%%%%%%%%%%%%%%%%%%%%%%%%%%%

%%%%%%%%%%%%%%%%%%%%%%%%%%%%%%%%%%%%%%%%%%%%%%%%%
\end{document}